\documentclass[prd,eqsecnum,twocolumn,amsfonts,amssymb]{revtex4}

\usepackage{graphicx}

\usepackage{bm}

\newcommand{\beq}{\begin{equation}}
\newcommand{\eeq}{\end{equation}}
\newcommand{\beqs}{\begin{eqnarray}}\newcommand{\eeqs}{\end{eqnarray}}
\newcommand{\lsim}{\mathrel{\raisebox{-
.6ex}{$\stackrel{\textstyle<}{\sim}$}}}
\newcommand{\gsim}{\mathrel{\raisebox{-
.6ex}{$\stackrel{\textstyle>}{\sim}$}}}

\newcommand{\drawsquare}[2]{\hbox{%
\rule{#2pt}{#1pt}\hskip-#2pt%  left vertical
\rule{#1pt}{#2pt}\hskip-#1pt%  lower horizontal
\rule[#1pt]{#1pt}{#2pt}}\rule[#1pt]{#2pt}{#2pt}\hskip-#2pt%  upper horizontal
\rule{#2pt}{#1pt}}% right vertical
\newcommand{\fund}{\raisebox{-.5pt}{\drawsquare{6.5}{0.4}}}%  fund
\newcommand{\sym}{\raisebox{-.5pt}{\drawsquare{6.5}{0.4}}\hskip-0.4pt%
        \raisebox{-.5pt}{\drawsquare{6.5}{0.4}}}%  symmetric second rank
\begin{document}

\title{Baryon-Number-Violating 
Nucleon and Dinucleon Decays in a Model with Large Extra Dimensions} 

\author{Sudhakantha Girmohanta and Robert Shrock}

\affiliation{ \ C. N. Yang Institute for Theoretical Physics and 
Department of Physics and Astronomy, \\
Stony Brook University, Stony Brook, NY 11794, USA }

\begin{abstract}

  It is known that limits on baryon-violating nucleon decays do not, in
  general, imply corresponding suppression of $n - \bar n$ transitions. In the
  context of a model with fermions propagating in higher dimensions, we
  investigate a related question, namely the implications of limits on $\Delta
  L=-1$ proton and bound neutron decays mediated by four-fermion operators for
  rates of nucleon decays mediated by $k$-fermion operators with $k =6$ and
  $k=8$. These include a variety of nucleon and dinucleon decays to dilepton
  and trilepton final states with $\Delta L=-3, \ -2, \ 1$, and $2$. We carry
  out a low-energy effective field theory analysis of relevant operators for
  these decays and show that, in this extra-dimensional model, the rates for
  these decays are strongly suppressed and hence are in accord with
  experimental limits.

\end{abstract}

\maketitle

% =======================================================================

% section 1
\section{Introduction}
\label{intro_section}

Although the Standard Model (SM), as extended to include nonzero neutrino
masses and lepton mixing, agrees with current data, there are many aspects of
particle physics that it does not explain. Although this theory conserves
baryon number, $B$ \ \cite{hooft}, many ultraviolet extensions of it predict
baryon number violation (BNV).  In general, one expects there to be some
violation of baryon number in nature, because this is one of the necessary
conditions for generating the observed baryon asymmetry in the universe
\cite{sakharov}.  A number of dedicated experiments have been carried out since
the early 1980s to search for baryon-number-violating decays of protons and of
neutrons bound in nuclei. (Henceforth, we shall refer to these as nucleon
decays, with it being understood that the term excludes
baryon-number-conserving weak decays of neutrons.) These experiments have
obtained null results and have set resultant stringent upper limits for the
rates of such nucleon decays \cite{pdg}.

It was pointed out early on that neutron-antineutron ($n-\bar n$) oscillations
and the associated $|\Delta B|=2$ violation of baryon number could account for
baryogenesis \cite{kuzmin}, and there has long been interest in this type of
baryon number violation (some early works include
\cite{earlynnbar}-\cite{nnb84}). The same physics beyond the Standard Model
(BSM) that gives rise to $n-\bar n$ oscillations also leads to matter
instability via the decays of $nn$ and $np$ dinucleon initial states to
nonbaryonic final states, typically involving several pions.  The reason for
this is that a nonzero transition amplitude $\langle \bar n |{\cal
  L}_{eff}|n\rangle$ means that a physical state $|n \rangle_{phys.}$ contains
a small but nonzero $|\bar n\rangle$ component. In turn, this leads to the
annihilation of the $|\bar n\rangle$ component with a neighboring neutron or
proton in a nucleus, and thus produces $\Delta B=-2$ decays of
dinucleons. There have been searches for $n-\bar n$ oscillations using neutron
beams from reactors \cite{ill} and for matter instability and various dinucleon
decay modes using large underground detectors
\cite{takita86}-\cite{sno_invisible}.

The operators in the low-energy effective Lagrangian for nucleon decay are
four-fermion operators with Maxwellian dimension 6 in mass units and hence
coefficients of the form $1/({\rm mass})^2$. In contrast, the operators in
${\cal L}^{(n \bar n)}_{eff}$ are six-quark operators with dimension 9 and
hence with coefficients of the form $1/({\rm mass})^5$.  Consequently, if one
were to assume that there is a single high mass scale $M_{BNV}$ describing the
physics responsible for baryon number violation, nucleon decay would be much
more important than $n-\bar n$ oscillations and the corresponding dinucleon
decays as a manifestation of baryon number violation.  However, the actual
situation might be quite different.  As was pointed out in Ref. \cite{nnb02}
and demonstrated explicitly using an extra-dimensional model \cite{as,ms},
nucleon decays could be suppressed well below an observable level, while
$n-\bar n $ oscillations could occur at a level comparable to existing
experimental limits.  In this case, it is the ($|\Delta B|=2$) $n-\bar n$
oscillations and the corresponding ($\Delta B = -2$) $nn$ and $np$ dinucleon
decays that are the main observable effects of baryon number violation, rather
than ($\Delta B = -1$) decays of individual nucleons.  Additional examples with
baryon number violation but no proton decay were later discussed in
\cite{wise}.  Reviews of $n-\bar n$ oscillations include
\cite{kamyshkov97,mohapatra_rev,nnbar_physrep}.

This finding in Ref. \cite{nnb02} naturally motivates one to ask a more general
question: in this type of extra-dimensional model, are there
baryon-number-violating processes mediated by $k$-fermion operators with higher
values of $k$, in particular, $k=6$ and $k=8$, that could also be relatively
unsuppressed, as was the case with the $k=6$ operators responsible for $n -
\bar n$ oscillations?

In this paper we address and answer this question. Using the same
extra-dimensional model as in \cite{nnb02}, we study a variety of nucleon and
dinucleon decays that violate both $B$ and total lepton number, $L$,
and are mediated by $k$-fermion operators with $k=6$ and
$k=8$, respectively.  These include the $\Delta L=-3$ nucleon decays
\beq
p \to \ell^+ \bar\nu \bar\nu'
\label{p_to_l2nubar}
\eeq
and
\beq
n \to \bar\nu \bar \nu' \bar \nu'' \ , 
\label{n_to_3nubar}
\eeq
and the $\Delta L=1$ nucleon decays 
\beq
p \to \ell^+ \nu \nu'
\label{p_to_l2nu}
\eeq
and 
\beq
n \to \bar\nu \nu' \nu'' \ , 
\label{n_to_nu2nubar}
\eeq
both of which are mediated by six-fermion operators, and the following 
$\Delta L=-2$ dinucleon decays mediated by eight-fermion operators: 
\beq
pp \to \ell^+ \ell'^+ \ , 
\label{pp_to_ll}
\eeq
where $\ell^+$ and $\ell'^+$ can be $e^+$, $\mu^+$, or $\tau^+$, as allowed by
phase space, i.e.,
\beq
pp \to (e^+e^+, \ \mu^+\mu^+, \ e^+\mu^+, \ e^+\tau^+, \ {\rm or} \ 
\mu^+\tau^+), 
\label{pp_to_ll_explicit}
\eeq
\beq
np \to \ell^+ \bar\nu \ , 
\label{np_to_lnubar}
\eeq
and 
\beq
nn \to \bar\nu \bar\nu' \ . 
\label{nn_to_2nubar}
\eeq
In addition, we consider the $\Delta L= 2$ dineutron decays
\beq
nn \to \nu \nu'  \ , 
\label{nn_to_2nu}
\eeq
which are also mediated by eight-fermion operators.  Here and below we use the
symbol $\nu$ to denote either an electroweak-doublet (EW-doublet) neutrino or
an EW-singlet neutrino.  From experimental limits on nucleon decays, we first
determine constraints on relevant parameters of the extra-dimensional model,
namely distances separating centers of fermion wavefunctions in the extra
dimensions.  Then, for each of the various types of decays, we analyze relevant
multi-fermion operators and apply these constraints to estimate the typical
predictions of the model for the decay rates.  Answering the question posed
above, we show that these nucleon decays
(\ref{p_to_l2nubar})-(\ref{n_to_nu2nubar}) and dinucleon decays
(\ref{pp_to_ll_explicit})-(\ref{nn_to_2nu}) are safely smaller than the rates
for the leading baryon-number-violating nucleon decays mediated by four-fermion
operators and thus are in accord with experimental limits.

There are several motivations for the class of extra-dimensional theories that
we consider.  The possibility that our four-dimensional spacetime could be
embedded in a higher-dimensional spacetime dates back at least to attempts to
unify electromagnetism and gravity by Kalusza and Klein \cite{kk}, and this
embedding is implied by string theory, since the low-energy limit of a
(super)string theory leads to a 10-dimensional pointlike field theory.  Since
all experimental data are consistent with spacetime being four-dimensional, the
extra dimensions must be compactified on scale(s) that is (are) much shorter
than those that have been probed experimentally.  In this context, the Standard
Model can be viewed as a low-energy effective field theory (EFT) that describes
physics at length scales much larger than the compactification scale(s).  One
of the most striking and perplexing features of the quarks and charged leptons
is the great range of approximately $10^5$ spanned by their masses, extending
from 173 GeV for the top quark to 0.511 MeV for the electron. The Standard
Model gives no insight into the reason for this large range of masses, and
instead just accommodates it via a correspondingly large range of magnitudes of
Yukawa couplings. This fermion mass hierarchy is even larger when one takes
into account the tiny but nonzero masses of neutrinos. An intriguing suggestion
was that this large range of SM fermion masses might be explained naturally if
the SM is embedded in a spacetime of higher dimension $d=4+n$, with $n$
extra additional spatial dimensions, and SM fermions have
wavefunctions that are localized at different positions in the additional
$n$-dimensional space \cite{as,ms}. Here we will use a model of this type in
which the wavefunctions of the SM fermions are strongly localized, with
Gaussian profiles of width $1/\mu$, at various points in this extra-dimensional
space \cite{nnb02}-\cite{ms}, \cite{trap}-\cite{c2}.  As in
Refs. \cite{as,ms,nnb02}, we do not make any specific assumption concerning 
possible ultraviolet completions of the model.

In addition to giving insight into various baryon- and lepton-number violating
processes in the context of a BSM model, our analysis is an interesting
application of effective field theory in a more complicated case than usual, in
which there are multiple mass scales relevant for the $B$ and $L$ violation,
namely $\mu$, a general scale $M_{BNV}$ characterizing baryon number
violation, and the inverse distances between the
centers of the wavefunctions of various fermions in the extra
dimensions.  For each decay with a given $\ell=e$ or $\mu$, there are at least
${6 \choose 2}=15$ of these inverse distances, corresponding to the five SM
quark and lepton fields $Q_L$, $u_R$, $d_R$, $L_{\ell,L}$, $\ell_R$, and one or
more electroweak-singlet neutrinos, $\nu_{s,R}$.  There is a correspondingly
large variety of multi-fermion operators with different structures, which we
analyze. 

The present work complements our recent studies in \cite{ndl}, where we
derived improved upper bounds on the rates for several nucleon-to-trilepton
decay modes with $\Delta L=-1$ and in \cite{dnd}, where we similarly presented
improved upper bounds on the rates for several dinucleon-to-dilepton decay
channels with $\Delta L=0$.  These works \cite{ndl,dnd} were model-independent 
phenomenological analyses, whereas our present paper is a study within the
context of a specific type of extra-dimensional model. 

This paper is organized as follows.  In Sec. \ref{background_section} we
discuss the extra-dimensional model and low-energy effective field theory
approach that serve as the theoretical framework for our calculations. In
Sec. \ref{pdecay_section} we extract constraints on the fermion wavefunctions
in the model from limits on nucleon decay modes.  Section \ref{nnbar_section}
is devoted to a review of $n - \bar n$ oscillations in the model, as mediated
by six-fermion operators.  A discussion is given in Sec. \ref{dnd_section} of
$\Delta L=0$ dinucleon decays to dileptons.  In Sects. \ref{nuc3_decay_section}
and \ref{nuc3b_decay_section} we analyze six-fermion operators that contribute
to $\Delta L=-3$ and $\Delta L=1$ nucleon decays to trilepton final states,
respectively. In Sec.  \ref{dinucleon_to_dilepton_section} we present a general
operator analysis of eight-fermion operators that contribute to $\Delta L= -2$
dinucleon decays to dileptons. Applications of this general analysis to the
decays $pp \to \ell^+ \ell'^+$, $np \to \ell^+ \bar\nu$, and $nn \to \bar\nu
\bar\nu'$ are given in Sections
\ref{pp_to_ll_section}-\ref{nn_to_2nubar_section}.  Section
\ref{nn_to_2nubar_section} also contains a discussion of the $\Delta L=2$
dineutron decays $nn \to \nu \nu'$.  Our conclusions are contained in Section
\ref{conclusion_section}. In Appendices \ref{integral_appendix},
\ref{colortensor_appendix}, and \ref{pp_operator_appendix} we give relevant
integral formulas, color SU(3)$_c$ and weak SU(2)$_L$ tensors, and present
further information on relevant operators.

% ========================================================================

\section{Theoretical Framework}
\label{background_section} 

In this section we describe the theoretical framework for our study.  Usual
spacetime coordinates are denoted as $x_\nu$, $\nu=0,1,2,3$, and the $n$
extra coordinates as $y_\lambda$; for definiteness, the latter are assumed to
be compact. The fermion fields are taken to have a factorized form,
\beq
\Psi(x,y)=\psi(x)\chi(y) \ .
\label{psiform}
\eeq
In the extra dimensions the SM fields are restricted to the 
interval $0 \le y_\lambda \le L$ for all $\lambda$.  
We define an energy corresponding to the inverse of the compactification scale
as 
\beq
\Lambda_L \equiv \frac{1}{L} \ . 
\label{lambda_L}
\eeq
We will give most results for general $n$, but note that only for even
$n$ are chiral projection operators defined, since they require there to be
a $\gamma_5$ Dirac matrix that anticommutes with the other Dirac gamma
matrices, and this is only possible for even $n$.  The
$d=(4+n)$-dimensional fields thus have Kaluza-Klein (KK) mode
decompositions.  We use a low-energy effective field theory approach that
entails an ultraviolet cutoff, which we denote as $M_*$. The localization of
the wavefunction of a fermion $f$ in the extra dimensions has the form
\cite{as,ms} 
\beq
\chi_f(y) = A \, e^{-\mu^2 \, \| y-y_f \|^2} \ , 
\label{gaussian}
\eeq
where $A$ is a normalization factor and $y_f \in {\mathbb R}^n$ denotes 
the position vector of this fermion in the extra dimensions, with components 
$y_f = ((y_f)_1,...,(y_f)_n)$ and with the standard Euclidean norm 
of a vector in ${\mathbb R}^n$, namely 
\beq
\| y_f \| \equiv \Big (\sum_{\lambda=1}^n y_{f,\lambda}^2 \Big )^{1/2} \ . 
\label{yfnorm}
\eeq
For $n=1$ or $n=2$, this fermion localization can result from appropriate
coupling to a scalar with a kink or vortex solution, respectively \cite{trap}.
One can also include corrections due to Coulombic gauge interactions between
fermions \cite{qlw} (see also \cite{gp,surujon}). The normalization factor 
$A$ is determined by the condition that, after integration over the $n$ 
higher dimensions, the four-dimensional fermion kinetic term has its canonical
normalization. This yields the result 
\beq
A=\bigg ( \frac{2}{\pi} \bigg )^{n/4}\, \mu^{n/2} \ . 
\label{a}
\eeq
We define a distance inverse to the localization measure $\mu$ as 
\beq
L_\mu \equiv \frac{1}{\mu} \ . 
\label{ellmu}
\eeq
As noted, this type of model has the potential to yield an explanation for the
hierarchy in the fermion mass matrices via the localization of fermion
wavefunctions with half-width
\beq
L_\mu \ll L
\label{lmul}
\eeq
at various points in the higher-dimensional space.  The ratio of the
compactification scale $L$ divided by the scale characterizing the localization
of the fermion wavefunctions in the extra dimensions is
\beq
\xi \equiv \frac{L}{L_\mu} = \frac{\mu}{\Lambda_L} = \mu L \ .
\label{xi}
\eeq
The choice
\beq
\xi \sim 30
\label{xivalue}
\eeq
is made for sufficient separation of the various fermion wavefunctions while
still fitting well within the size $L$ of the compactified extra
dimensions. The UV cutoff $M_*$ satisfies $M_* > \mu$ for the validity the
low-energy field theory analysis. The choice
\beq
\Lambda_L \gsim 100 \ {\rm TeV} \ , 
\label{lambda_value}
\eeq
i.e., $L \lsim 2.0 \times 10^{-19}$ cm, is consistent with bounds on extra
dimensions from precision electroweak constraints, and collider searches
\cite{pdg} and produces adequate suppression of flavor-changing neutral-current
(FCNC) processes \cite{c1,c2}.  With the ratio $\xi=30$, this yields
\beq
\mu \sim 3 \times 10^3 \ {\rm TeV} \ , 
\label{muvalue}
\eeq
i.e., $L_\mu \equiv \mu^{-1} = 0.67 \times 10^{-20}$ cm. 

Starting from an effective Lagrangian in the $d=(4+n)$-dimensional spacetime,
one obtains the resultant low-energy effective Lagrangian in four dimensions by
integrating over the extra $n$ dimensions.  The integration over each of the
$n$ coordinates of a vector $y$ runs from 0 to $L$, but, because of the
restriction of the fermion wavefunctions to the form (\ref{gaussian}), with
$L_\mu \ll L$, it follows that, to a very good approximation, the domain of
integration can be extended to the interval $(-\infty,\infty)$: $\int_0^L
d^n y \to \int_{-\infty}^\infty d^n y$.  It is convenient to define the
dimensionless variable
\beq
\eta = \mu y \ , 
\label{eta}
\eeq
with components given by $\eta = (\eta_1,...,\eta_n)$.  

We first discuss the fermion mass terms. For the first generation of quarks and
charged leptons, the Yukawa terms in the higher-dimension theory are 
\beqs
{\cal L}_{Yuk} &=& \Big [ h^{(d)} \bar Q_L d_R \phi + 
                 h^{(u)} \bar Q_L u_R \tilde\phi + 
                 h^{(e)} \bar L_{e,L} e_R \phi \Big ] \cr\cr
               &+& h.c. \ ,
\label{yukterm}
\eeqs
where $Q_L = {u \choose d}_L$, and $\phi = {\phi^+ \choose \phi^0}$ is the SM
Higgs field, with $\tilde \phi = i \sigma_2 \phi^\dagger = {\phi^{0 *} \choose
  -\phi^-}$. With the inclusion of the second and third generations of SM
fermions, the Yukawa couplings $h^{(f)}$ with $f=u, \ d, \ e$ become $3 \times
3$ matrices. The diagonalization of the resultant quark mass matrices in the
charge 2/3 and charge $-1/3$ sectors yields the quark masses and
Cabibbo-Kobayashi-Maskawa quark mixing matrix. For our present purposes,
it will often be adequate to neglect small off-diagonal elements in the Yukawa
matrices. The vacuum expectation value of the Higgs field is written, in
the standard normalization, as
\beq
\langle \phi \rangle_0 = {0 \choose v/\sqrt{2}} \ ,
\label{higgs}
\eeq
where $v=246$ GeV.  Given the factorization (\ref{psiform}) and the Gaussian
profiles of the fermion wave functions (\ref{gaussian}), the integration over
the extra $n$ dimensions of a given fermion bilinear operator product
$h^{(f)}(v/\sqrt{2}) [\bar f_L f_R]$ resulting from a Yukawa interaction
involves the integral
\beqs
&& A^2 \, h^{(f)} \, \frac{v}{\sqrt{2}} \, 
\int d^n y \, e^{-\|\eta-\eta_{f_L}\|^2- \|\eta-\eta_{f_R}\|^2}
\cr\cr 
&=& h^{(f)} \, \frac{v}{\sqrt{2}} \exp \Big [ 
-\frac{1}{2}\|\eta_{f_L} - \eta_{f_R}\|^2 \Big ] \ . 
\label{mint}
\eeqs
Hence, for the fermions $f=u, \ d$ and also $f=\ell=e, \ , \mu, \ \tau$
(neglecting off-diagonal elements in the Yukawa matrices), we have 
\beq
m_f = h^{(f)} \, \frac{v}{\sqrt{2}} \exp \Big [ 
-\frac{1}{2}\|\eta_{f_L} - \eta_{f_R}\|^2 \ \Big ] \ , 
\label{mf}
\eeq
or equivalently, the following constraint on the separation distance 
$\|\eta_{f_L} - \eta_{f_R}\|$:
\beq
\|\eta_{f_L} - \eta_{f_R}\| = \bigg [ 2\ln\bigg ( 
\frac{h^{(f)}v}{\sqrt{2} \, m_f} \bigg ) \bigg ]^{1/2} \ . 
\label{mf_distance_constraint}
\eeq
Note that this relation does not depend directly on the number of large extra
dimensions, $n$. The relation (\ref{mf_distance_constraint}) holds for the
quarks and charged leptons.  For neutrinos, the situation is more complicated
because the neutrino mass eigenvalues and the lepton mixing matrix result, in
general, from the diagonalization of the combined Dirac and Majorana mass terms
involving electroweak-singlet neutrinos $\nu_{s,R}$, $s=1,...,n_s$. These
Majorana neutrino mass terms violate $L$ (as $|\Delta L|=2$ operators) and lead
to potentially observable $L$-violating processes. However, $L$-violation can
occur even with very small neutrino masses, as in $R$-parity-violating
supersymmetric theories (e.g., \cite{ls2000}).

Since the relation (\ref{mf}) applies in the effective Lagrangian above the
electroweak-symmetry-breaking scale, the values of $m_f$ are the
running masses evaluated at this high scale. In accord with the idea
motivating this class of BSM theories, that the generational hierarchy in the
SM fermion masses is not due primarily to a hierarchy in the dimensionless
Yukawa couplings in the higher-dimensional space, but instead to the different
positions of the wavefunction centers in the extra dimensions, we will take
$h^{(f)} \sim O(1)$ in the higher-dimensional space for the various SM fermions
$f$.  For technical simplicity, we actually set $h^{(f)} = 1$ for all $f$.  It
is straightfoward to redo our analysis if one chooses to assign some of the
generational mass hierarchy to these Yukawa couplings in the
$(4+n)$-dimensional space.  A calculation of the running quark masses at a
scale $\Lambda_t=m_t$ gives \cite{koide} $m_u(\Lambda_t)=2.2$ MeV and
$m_d(\Lambda_t)=4.5$ MeV.  Combining these these values with the known value
$v=246$ GeV from $G_F/\sqrt{2} = 1/(2v^2)$, we calculate the dimensionless
separation distances
\beq
\|\eta_{Q_L}-\eta_{u_R}\|=4.75
\label{distance_QL_ur}
\eeq
and
\beq
\|\eta_{Q_L}-\eta_{d_R}\|=4.60 \ , 
\label{distance_QL_dr}
\eeq
so that the ratio is $\|\eta_{Q_L}-\eta_{u_R}\|/
\|\eta_{Q_L}-\eta_{d_R}\|=1.03$ \cite{compare_ms}.

As noted, a major result from this type of model was the fact that with roughly
equal dimensionless Yukawa couplings $h^{(f)} \sim O(1)$ for different
generations of quark and charged leptons, the large hierarchy in the values of
these SM fermion masses can be explained by moderate differences in the
separation distances in the extra dimensions, $\|\eta_{f_L} - \eta_{f_R}\|$.
This extra-dimensional model is minimal in the sense that we do not include
additional fields aside from neutrinos that carry lepton number, such as
Majorons.

A given baryon-number-violating decay involves a set of operators defined in
four-dimensional spacetime, which, for our applications, are $k$-fold products 
of fermion fields.  We denote these operators as ${\cal O}_{r,(k)}$ and write 
the effective Lagrangian in usual four-dimensional spacetime that
is responsible for the BNV physics as
\beq
{\cal L}_{eff}(x) = \sum_r c_{r,(k)} {\cal O}_{r,(k)}(x) + h.c. \ , 
\label{leff}
\eeq
Each of the fermion fields in ${\cal O}_{r,(k)}$ has the factorized form
(\ref{psiform}).  We denote the corresponding effective Lagrangian in the
$d=(4+n)$-dimensional space as 
\beq
{\cal L}_{eff,4+n}(x,y) = \sum_r \kappa_{r,(k)} O_{r,(k)}(x,y) + h.c. \ . 
\label{leff_higherdim}
\eeq
The factorization property (\ref{psiform}) implies that the $O_{r,(k)}(x,y)$ 
also can be factored as 
\beq
O_{r,(k)}(x) = U_{r,(k)}(x) V_{r,(k)}(y) 
\label{uvfactorization}
\eeq
(with SU(3)$_c$, SU(2)$_L$, and Dirac structure implicit and 
with no sum on $r$). We denote the integral over the extra dimensions of
$V_{r,(k)}(y)$ as
\beq
I_{r,(k)} \equiv \int d^n y \, V_{r,(k)}(y) \ . 
\label{integral_r}
\eeq
This integral involves an integrand consisting of a $k$-fold product of 
Gaussian wavefunctions and is given by Eq. (\ref{intform}) in Appendix 
\ref{integral_appendix}.  Hence, for each $r$ (with no sum on $r$) 
\beq
c_{r,(k)} = \kappa_{r,(k)} I_{r,(k)} \ . 
\label{crk}
\eeq
The coefficient $\kappa_{r,(k)}$ may depend on the generational indices of
lepton fields that occur in ${\cal O}_{r,(k)}$; this is left implicit in the
notation. In general, as a $k$-fold product of fermion fields in $d=4+n$
spacetime dimensions, $O_{r,(k)}(x,y)$ has Maxwellian (free-field) operator
dimension
\beq
{\rm dim}(O_{r,(k)}(x,y)) = \frac{k(d-1)}{2} = \frac{k(3+n)}{2} 
\label{dim_op}
\eeq
in mass units.  The condition that the action in the
$d$-dimensional space must be dimensionless is
$-d+{\rm dim}(\kappa_{r,(k)}) + {\rm   dim}(O_{r,(k)}) = 0$, so
\beqs
{\rm dim}(\kappa_{r,(k)}) &=& d-k\Big ( \frac{d-1}{2} \Big ) \cr\cr
                    &=& 4+n - k \Big ( \frac{3+n}{2} \Big ) \ . 
\label{dim_kappa}
\eeqs
It is useful to write the coefficients
$\kappa_{r,(k)}$ in a form that shows this dimensionality explicitly:
\beq
\kappa_{r,(k)} = \frac{\bar\kappa_{r,(k)}}
{(M_{BNV})^{(k(3+n)/2)-4-n}} \ , 
\label{kappagen}
\eeq
where $\bar\kappa_{r,(k)}$ is dimensionless and $M_{BNV}$ is an effective 
mass scale characterizing the baryon-number violating physics. Then, making use
of Eq. (\ref{intform}), $I_{r,(k)}$ can be written as a prefactor $b_k$
multiplying an exponential, namely 
\beq
I_{r,(k)} = b_k \, e^{-S_{r,(k)}} \ , 
\label{irgen}
\eeq
where 
\beqs
&& b_k = A^k \, \mu^{-n}\Big ( \frac{\pi}{k} \Big )^{n/2} \cr\cr
&&=\Big [ 2^{k/4} \, \pi^{-(k-2)/4} \, k^{-1/2} \, \mu^{(k-2)/2} \Big ]^n \
.
\label{bk}
\eeqs
In Eq. (\ref{bk}), the factor $A^k$ arises from the $k$-fold product of fermion
fields, the factor $\mu^{-n}$ from the Jacobian $d^n y = \mu^{-n} \,
d^n \eta$, and the factor $(\pi/k)^{n/2}$ arises from the integration
(see Eq. (\ref{intform}) in Appendix \ref{integral_appendix}).  By 
construction, $b_2=1$, independent of the number of large extra dimensions,
$n$.  Combining these results, we can write 
\begin{widetext}
\beq
c_{r,(k)} = \kappa_{r,(k)} I_{r,(k)} 
 =  \frac{\bar\kappa_{r,(k)}}{(M_{BNV})^{(3k-8)/2} } \, 
\Big ( \frac{\mu}{M_{BNV}} \Big ) ^{(k-2)n/2} \, 
\bigg ( \frac{2^{k/4}}{\pi^{(k-2)/4} \, k^{1/2} } \bigg )^n \,
e^{-S_{r,(k)}}
\ . 
\label{crgen}
\eeq
\end{widetext}
For each of the various types of decays discussed below, the number $k$ of
fermions in the $k$-fermion operator products will be obvious, so henceforth,
we suppress the subscript $(k)$ in the notation for $I_{r,(k)}$ and $c_{r,(k)}$.

Before carrying out detailed analyses of various baryon-number-violating
decays, it is useful to make some rough estimates of the expected ratios of
resultant rates.  The hadronic matrix elements that are relevant for decays
mediated by operators with different numbers of fermions
have different dimensions, but in comparing decay rates, this difference is
compensated by the requisite powers of the quantum chromodynamics (QCD) mass
scale, $\Lambda_{QCD}$.  Thus, for the ratio of two BNV decays mediated by
operators comprised of $k_1$ and $k_2$ fermions, respectively, we have the rough
estimate
\begin{widetext}
\beq
 \frac{\Gamma_{(k_2)}}{\Gamma_{(k_1)}} \sim
\Big ( \frac{\Lambda_{QCD}}{M_{BNV}} \Big )^{3(k_2-k_1)} \, 
\Big ( \frac{\mu}{M_{BNV}} \Big )^{(k_2-k_1)n} \, 
\bigg ( \frac{2^{(k_2-k_1)/2} k_1}{\pi^{(k_2-k_1)/2} k_2} \bigg )^n \, 
 e^{-2(\langle S_{(k_2)}\rangle - \langle S_{(k_1)}\rangle ) } \ , 
\label{ratio_k21}
\eeq
\end{widetext}
where $\langle S_{(k)}\rangle$ denotes a typical size of the exponential factor
occurring in Eqs. (\ref{irgen}) and (\ref{crgen}) for this decay. 
In particular, relative to BNV nucleon decays such as $p \to e^+ \pi^0$,
etc. mediated by four-fermion operators, the rough estimate 
(\ref{ratio_k21}) gives the ratio 
\beqs
 \frac{\Gamma_{(6)}}
      {\Gamma_{(4)}} &\sim&
\Big ( \frac{\Lambda_{QCD}}{M_{BNV}} \Big )^6 \, 
\Big ( \frac{\mu          }{M_{BNV}} \Big )^{2n} \,
\Big ( \frac{4}{3\pi} \Big )^n \, 
e^{-2(\langle S_{(6)}\rangle - 
      \langle S_{(4)}\rangle ) } \cr\cr
&&
\label{ratio_k64}
\eeqs
for decays such as (\ref{p_to_l2nubar})-(\ref{n_to_nu2nubar}) mediated by
six-fermion operators.  Similarly, for dinucleon decays mediated by
eight-fermion operators, such as (\ref{pp_to_ll_explicit})-(\ref{nn_to_2nu}),
Eq. (\ref{ratio_k21}) predicts
\beqs
 \frac{\Gamma_{(8)}}
      {\Gamma_{(4)}} &\sim&
\Big ( \frac{\Lambda_{QCD}}{M_{BNV}} \Big )^{12} \, 
\Big ( \frac{\mu}{M_{BNV}} \Big )^{4n} \,
\Big ( \frac{2}{\pi^2} \Big )^n \, 
e^{-2(\langle S_{(8)}\rangle - 
      \langle S_{(4)}\rangle ) } \ . \cr\cr
&&
\label{ratio_k84}
\eeqs
With $\Lambda_{QCD} \simeq 0.25$ GeV, $\mu = 3 \times 10^3$ TeV 
as in Eq. (\ref{muvalue}), and an illustrative values $M_{BNV} \sim 10^2$ TeV
and $n=2$ extra dimensions, Eqs. (\ref{ratio_k64}) and (\ref{ratio_k84})
yield 
\beq
\ln \bigg ( \frac{\Gamma_{(6)}}
 {\Gamma_{(4)}} \bigg ) \sim -65.5 
- 2(\langle S_{(6)}\rangle - \langle S_{(4)}\rangle ) 
\label{log_ratio_k64_example}
\eeq
and
\beq
\ln \bigg ( \frac{\Gamma_{(8)}}
 {\Gamma_{(4)}} \bigg ) \sim -131 
-2(\langle S_{(8)}\rangle - \langle S_{(4)}\rangle ) \ . 
\label{log_ratio_k84_example}
\eeq
The study of the sums $S_{r,(k)}$ requires a detailed analysis of the
various $k$-fermion operators that contribute to specific
baryon-number-violating processes. We discuss these below. 

% =========================================================================

\section{Constraints from Limits on Baryon-Number-Violating Nucleon Decays} 
\label{pdecay_section}

We discuss here the constraints on Standard-Model fermion wavefunction
positions in the extra-dimensional model that follow from the upper limits on
the rates for baryon-number-violating nucleon decays. The analysis begins with
the observation that the mass scale characterizing the physics responsible for
these decays must be large compared with the electroweak symmetry-breaking
scale, $v$, and therefore the effective Lagrangian must be invariant
under the full Standard-Model gauge group, $G_{SM}$.  To label the various
(four-fermion) operators that contribute, we will use the abbreviations $pd$
and $nd$ to refer to proton and (otherwise stably bound) neutron decay and $Nd$
to subsume both of these types of decay, with the nucleon $N=p$ or $N=n$. (The
use of the same symbol, $n$, to refer to neutron and the number of extra
dimensions should not cause any confusion; the context will always make clear
which is meant.) Then we can write
\beq
{\cal L}^{(Nd)}_{eff}(x) = \sum_r c^{(Nd)}_r {\cal O}^{(Nd)}_r(x) + h.c. \ , 
\label{leff_pd}
\eeq
where $c^{(Nd)}_r$ are coefficients, and ${\cal O}^{(Nd)}_r(x)$ are operators.
Correspondingly, in the $d=(4+n)$-dimensional space, the effective Lagrangian
is
\beq
{\cal L}^{(Nd)}_{eff,4+n}(x,y) = 
\sum_r \kappa^{(Nd)}_r O^{(Nd)}_r(x,y) + h.c. \ . 
\label{leff_higherdim_pd}
\eeq

We recall our notation for fermion fields. 
The SU(2)$_L$-singlet and doublet quark fields are denoted 
$u^\alpha_R$, $d^\alpha_R$, and $Q^\alpha_L = {u^\alpha \choose d^\alpha}_L$, 
where $\alpha$ is a color index.  The SU(2)$_L$-singlet and
SU(2)$_L$-doublet lepton fields are denoted $\ell_R$ and $L_{\ell,L} = 
{\nu_\ell \choose \ell}_L$, where $\ell=e, \ \mu, \tau$. In addition, we 
include electroweak-singlet neutrinos, written as
$\nu_{s,R}$, with $s=1,...,n_s$, as is necessary to form Dirac and Majorana
mass terms for the neutrinos. 
The upper and lower components of the quark and
lepton SU(2)$_L$ doublets are indicated by Roman indices $i,j,..$, so 
$Q^{i \alpha}_L=u^\alpha_L$ for $i=1$, 
$Q^{i \alpha}_L=d^\alpha_L$ for $i=2$, 
$L^i_{\ell,L} = \nu_\ell$ for $i=1$, and 
$L^i_{\ell,L} = \ell_L$ for $i=2$. 
For each of these fields $f=Q_L, \ u_R, \ d_R, L_{\ell,L}$, $\ell_R$, and
$\nu_{s,R}$, the wavefunction in the $(4+n)$-dimensional space has the form
(\ref{psiform}) with normalization factor $A$ given by Eq. (\ref{a}) and
Gaussian profile given by Eq. (\ref{gaussian}).

With the original SM fermions, before the addition of any electroweak-singlet
$\nu_{s,R}$ fields, the four-fermion operators ${\cal O}^{(Nd)}_r$ in
${\cal L}^{(Nd)}_{eff}$ that contribute to nucleon decays
are \cite{weinberg79}-\cite{abbott_wise}
\beq
{\cal O}^{(Nd)}_1 = \epsilon_{\alpha\beta\gamma}
[u^{\alpha \ T}_R C d^\beta_R][u^{\gamma \ T}_R C \ell_R]
\label{op1_nucdec}
\eeq
\beqs
{\cal O}^{(Nd)}_2 &=&  \epsilon_{ij} \epsilon_{\alpha\beta\gamma}
[Q^{i \alpha \ T}_L C Q^{j \beta}_L][u^{\gamma \ T}_R C \ell_R] \cr\cr
&=& 2\epsilon_{\alpha\beta\gamma}
[u^{\alpha \ T}_L C d^\beta_L][u^{\gamma \ T}_R C \ell_R]
\label{op2_nucdec}
\eeqs
\beqs
&& {\cal O}^{(Nd)}_3 = \epsilon_{ij}\epsilon_{\alpha\beta\gamma}
[Q^{i \alpha \ T}_L C L^j_{\ell,L}][u^{\beta \ T}_R C d^\gamma_R] \cr\cr
&& = \epsilon_{\alpha\beta\gamma}\Big ( [u^{\alpha \ T}_L C \ell_L] -
       [d^{\alpha \ T}_L C \nu_{\ell,L}] \Big )[u^{\beta \ T}_L C d^\gamma_R]
 \cr\cr
&&
\label{op3_nucdec}
\eeqs
and
\beqs
{\cal O}^{(Nd)}_4 &=&
\epsilon_{ij} \epsilon_{km} \epsilon_{\alpha\beta\gamma}
[Q^{i \alpha \ T}_L C Q^{j \beta}_L][Q^{k \gamma \ T}_L C L^m_{\ell,L}] 
\cr\cr
&=& 2\epsilon_{\alpha\beta\gamma}[u^{\alpha \ T}_L C d^\beta_L]
\Big ( [u^{\gamma \ T}_L C \ell_L] -
       [d^{\gamma \ T}_L C \nu_{\ell,L}] \Big ) \ , \cr\cr
&&
\label{op4_nucdec}
\eeqs
where $C$ is the Dirac charge conjugation matrix satisfying $C \gamma_\mu
C^{-1} = -(\gamma_\mu)^T$, $C=-C^T$; and $\epsilon_{\alpha\beta\gamma}$ and
$\epsilon_{ij}$ are totally antisymmetric SU(3)$_c$ and SU(2)$_L$ tensors,
respectively. Two other operators would be present in a multigenerational
context but vanish identically in the relevant case here, where the quarks are
all of the first generation, i.e., $u$ and $d$:
\beq
\epsilon_{\alpha\beta\gamma}
[u^{\alpha \ T}_{a_1,R} C u^\beta_{a_2,R}][d^{\gamma \ T}_{a_3,R} C
\ell_R]
\label{opzero1_nucdec}
\eeq
and
\beqs
&&(\epsilon_{ik}\epsilon_{jm} + \epsilon_{im}\epsilon_{jk})
\epsilon_{\alpha\beta\gamma} \times \cr\cr
&\times&[Q^{i \alpha \ T}_{a_1,L} C Q^{j \beta}_{a_2,L}]
  [Q^{k \gamma \ T}_{a_3,L} C L^m_{\ell,L}] \ , 
\label{opzero2_nucdec}
\eeqs
where $a_1$, $a_2$, and $a_3$ are generation indices. 

Including electroweak-singlet neutrinos $\nu_{s,R}$ with $s=1,...,n_s$,
one has two additional types of operators for nucleon decays, namely
\beq
{\cal O}^{(Nd)}_5 = \epsilon_{\alpha\beta\gamma}
[u^{\alpha \ T}_R C d^\beta_R][d^{\gamma \ T}_R C \nu_{s,R}]
\label{op5_nucdec}
\eeq
and
\beqs
{\cal O}^{(Nd)}_6 &=& \epsilon_{ij}\epsilon_{\alpha\beta\gamma}
[Q^{i \alpha \ T}_L C Q^{j \beta}_L][d^{\gamma \ T}_R C \nu_{s,R}] \cr\cr
&=& 2\epsilon_{\alpha\beta\gamma}
[u^{\alpha \ T}_L C d^\beta_L][d^{\gamma \ T}_R C \nu_{s,R}] \ .
\label{op6_nucdec}
\eeqs
For completeness, we also list a four-fermion operator that would be present in
a multigenerational context but vanishes identically in the case considered
here with first-generation quarks, namely
\beq
\epsilon_{\alpha\beta\gamma}
[d^{\alpha \ T}_{a_1,R} C d^\beta_{a_2,R}]
[u^{\gamma \ T}_{a_3,R} C \nu_{s,R}] \ .
\label{opzero3_nucdec}
\eeq

To each of the operators ${\cal O}^{(Nd)}_r$ there corresponds an operator
$O^{(Nd)}_r$ in ${\cal L}^{(Nd)}_{eff,4+n}$. These are four-fermion
operators, and, as the $k=4$ special case of Eq. (\ref{kappagen}), we have
\beq
\kappa^{(Nd)}_r = \frac{\bar\kappa^{(Nd)}_r}{(M_{BNV})^{2+n}} \ , 
\label{kappabarpd}
\eeq
As noted before, in general,
the coefficient $\kappa^{(Nd)}_r$ may depend on the generational indices of
fermion fields that occur in ${\cal O}^{(Nd)}_r$; this is left implicit in the
notation. The special case of Eq. (\ref{uvfactorization}) for nucleon decay is 
\beq
O^{(Nd)}_r(x,y) = U^{(Nd)}_r(x)V^{(Nd)}_r(y) \ . 
\label{opd_st}
\eeq

We have 
\begin{widetext}
\beq
V^{(Nd)}_1(y) = A^4 \exp \Big [ - \Big \{ 
2\|\eta-\eta_{u_R}\|^2 +  
 \|\eta-\eta_{d_R}\|^2 + 
 \|\eta-\eta_{\ell_R} \|^2 \Big \} \Big ] 
\label{chiprod_op1_nucdec}
\eeq
\beq
V^{(Nd)}_2(y) = A^4 \exp \Big [ - \Big \{ 
2\|\eta-\eta_{Q_L}\|^2 +
 \|\eta-\eta_{u_R}\|^2 +
 \|\eta-\eta_{\ell_R} \|^2 \Big \} \Big ] 
\label{chiprod_op2_nucdec}
\eeq
\beq
V^{(Nd)}_3(y) = A^4 \exp \Big [ - \Big \{ 
 \|\eta-\eta_{Q_L}\|^2 + 
 \|\eta-\eta_{L_{\ell,L}}\|^2 + 
 \|\eta-\eta_{u_R}\|^2 + 
 \|\eta-\eta_{d_R}\|^2 \Big \} \Big ] 
\label{chiprod_op3_nucdec}
\eeq
\beqs
V^{(Nd)}_4(y) = A^4 \exp \Big [ - \Big \{ 
3\|\eta-\eta_{Q_L}\|^2 + 
 \|\eta-\eta_{L_{\ell,L}}\|^2 \Big \} \Big ]  
\label{chiprod_op4_nucdec}
\eeqs
\beqs
V^{(Nd)}_5(y) = A^4 \exp \Big [ - \Big \{
 \|\eta-\eta_{u_R}\|^2 + 
2\|\eta-\eta_{d_R}\|^2 +
 \|\eta-\eta_{\nu_{s,R}} \|^2 \Big \} \Big ] 
\label{chiprod_op7_nucleondecay}
\eeqs
and
\beq
V^{(Nd)}_6(y) = A^4 \exp \Big [ - \Big \{ 
2\|\eta-\eta_{Q_L}\|^2 + 
 \|\eta-\eta_{d_R}\|^2 + 
 \|\eta-\eta_{\nu_{s,R}} \|^2 \Big \} \Big ] \ . 
\label{chiprod_op8_nucdec}
\eeq

To perform the integrals over $y$, we use the general integration formula given
as Eq. (\ref{intform}) in Appendix \ref{integral_appendix}. Carrying out the
integration over the $y$ components and using Eq. (\ref{a}) for the relevant
case $k=4$, we obtain the following results for the nonvanishing operators:
\beq
I^{(Nd)}_1 = b_4 \, \exp \bigg [ -\frac{1}{4}\Big \{ 
2\|\eta_{u_R}-\eta_{d_R}\|^2 + 
2\|\eta_{u_R}-\eta_{\ell_R}\|^2 + 
 \|\eta_{d_R}-\eta_{\ell_R}\|^2 \Big \} \bigg ] 
\label{int_op1_nucdec}
\eeq
\beq
I^{(Nd)}_2 = b_4 \, \exp \bigg [ -\frac{1}{4} \Big \{ 
2\|\eta_{Q_L}-\eta_{u_R}\|^2 + 
2\|\eta_{Q_L}-\eta_{\ell_R}\|^2 + 
 \|\eta_{u_R}-\eta_{\ell_R}\|^2 \Big \} \bigg ] 
\label{int_op2_nucdec}
\eeq
\beqs
I^{(Nd)}_3 &=& b_4 \, \exp \bigg [ -\frac{1}{4}\Big \{
\|\eta_{Q_L}-\eta_{L_{\ell,L}}\|^2 + 
\|\eta_{Q_L}-\eta_{u_R}\|^2 + 
\|\eta_{Q_L}-\eta_{d_R}\|^2 + \cr\cr
&+& 
\|\eta_{L_{\ell,L}}-\eta_{u_R}\|^2 + 
\|\eta_{L_{\ell,L}}-\eta_{d_R}\|^2 + 
\|\eta_{u_R}-\eta_{d_R}\|^2 \Big \} \bigg ] 
\label{int_op3_nucdec}
\eeqs
\beq
I^{(Nd)}_4 = b_4 \, \exp \bigg [ 
-\frac{3}{4}\|\eta_{Q_L}-\eta_{L_{\ell,L}}\|^2 \bigg ] 
\label{int_op4_nucdec}
\eeq
\beq
I^{(Nd)}_5 = b_4 \, \exp \bigg [ -\frac{1}{4}\Big \{ 
2\|\eta_{u_R}-\eta_{d_R}\|^2 + 
 \|\eta_{u_R}-\eta_{\nu_{s,R}}\|^2 + 
2\|\eta_{d_R}-\eta_{\nu_{s,R}} \|^2 \Big \} \bigg ] 
\label{int_op5_nucdec}
\eeq
and
\beq
I^{(Nd)}_6 = b_4 \, \exp \bigg [ -\frac{1}{4}\Big \{ 
2\|\eta_{Q_L}-\eta_{d_R}\|^2 + 
2\|\eta_{Q_L}-\eta_{\nu_{s,R}}\|^2 + 
 \|\eta_{d_R}-\eta_{\nu_{s,R}} \|^2 \Big \} \bigg ] \ , 
\label{int_op6_nucdec}
\eeq
\end{widetext} 
where $b_4 = (\pi^{-1/2} \mu )^n$, from the $k=4$ special case of
Eq. (\ref{bk}). It is convenient to write the integral $I^{(Nd)}_r$ in the 
form 
\beq
I^{(Nd)}_r \equiv b_4 \, e^{-S^{(Nd)}_r} \ , 
\label{bsi}
\eeq
where $S^{(Nd)}_r$ denotes the sum of squares of fermion wavefunction
separation distances (rescaled via multiplication by $\mu$ to be dimensionless)
in the argument of the exponent in $I^{(Nd)}_r$. Thus, for example, in
the case of $O^{(Nd)}_4$, the sum in the exponent is 
$S^{(Nd)}_4 = (3/4)\|\eta_{Q_L}-\eta_{L_{\ell,L}}\|^2$, and similarly for the
other $S^{(Nd)}_r$.  Then, as the $k=4$ special case of (\ref{crgen}), 
\begin{widetext}
\beq
c^{(Nd)}_r = \frac{\bar\kappa^{(Nd)}_r}{(M_{BNV})^2} \, 
\bigg ( \frac{\mu}{\pi^{1/2} M_{BNV}} \bigg )^n \,  e^{-S^{(Nd)}_r} \ . 
\label{cr_nucleondecay}
\eeq
\end{widetext}

The amplitude for the decay of a nucleon $N=p$ or $n$ to a final state $f.s.$
is given by $\langle f.s.|{\cal O}_{eff}^{(Nd)}|N\rangle$.  The hadronic matrix
elements for various operators have been calculated by lattice gauge
simulations \cite{saoki_pdec,yaoki_pdec}. We then use the experimental lower
bound for the partial lifetime $(\tau/B)_{N \to f.s.} = \Gamma_{N \to
  f.s.}^{-1}$ for a given nucleon decay mode $N \to f.s.$ with branching ratio
$B$ to infer upper bounds on the magnitudes of the $c^{(Nd)}_r$
coefficients. Since in our low-energy effective field theory approach, we do
not assume any cancellation between different terms $c^{(Nd)}_r {\cal
  O}^{(Nd)}_r$ occurring in ${\cal L}^{(Nd)}_{eff}$, we conservatively impose
the bounds from a given decay individually on each term that contributes to it.
For given values of $\mu$, $M_{BNV}$, and the dimensionless coefficients
$\bar\kappa^{(Nd)}_r$, these constraints are upper bounds on the integrals
$I^{(Nd)}_r$
\beq
I^{(nd)}_r < I^{(Nd)}_{\rm max} \ , 
\label{irmax}
\eeq
and hence lower bounds on the the sums of squares of
distances in $S^{(Nd)}_r$ for each operator ${\cal O}^{(Nd)}_r$, 
\beq
S_r^{(Nd)} > S^{(Nd)}_{ \rm min} \ , 
\label{srbound}
\eeq
where 
\beq
S^{(Nd)}_{\rm min} = \ln \bigg ( \frac{b_4}{I^{(Nd)}_{\rm max}} \bigg ) \ . 
\label{srmin}
\eeq
When comparing lower bounds from two different nucleon decay modes, denoted
$Nd1$ and $Nd2$, to which the same operators contribute, a general relation is
\beq
S^{(Nd)}_{\rm min,1} - S^{(Nd)}_{\rm min,2} = \frac{1}{2} \ln \bigg [
\frac{(\tau/B)_{Nd1,{\rm min}}}{(\tau/B)_{Nd2,{\rm min}}} \bigg ] \ . 
\label{sdif}
\eeq
Some of the squared fermion separation distances $\|\eta_{f_i} -
\eta_{f_j}\|^2$ occurring in the individual $S^{(Nd)}_r$ sums are already fixed
by Standard-Model physics such as quark and lepton masses and mixing, and
values of, or limits on, flavor-changing neutral-current (FCNC)
processes. These include the (rescaled) distances $\|\eta_{Q_L}-\eta_{q_R}\|$
with $q_R=u_R, \ d_R$, and for leptons, the distances
$\|\eta_{L_{\ell,L}}-\ell'_R\|$ with $L_{\ell,L}=L_{e,L}, \ L_{\mu,L}, \
L_{\tau,L}$ and $\ell'_R=e_R, \ \mu_R, \ \tau_R$, respectively. For example,
for $\ell=e$, the inequality $S^{(Nd)}_2 > S^{(Nd)}_{\rm min}$ is a quadratic
inequality in the space ${\mathbb R}^{3n}$ spanned by the three $n$-dimensional
vectors $\eta_{Q_L}$, $\eta_{u_R}$, and $\eta_{\ell_R}$, with one distance
$\|\eta_{Q_L}-\eta_{u_R}\|$ fixed by the $u$-quark mass.  The (rescaled)
separation distances between SM fermion wavefunction centers that enter into
the $S^{(Nd)}_r$ of this type and are not already fixed by SM physics are
\beqs
&&
\|\eta_{u_R}-\eta_{d_R}\|, \quad 
\|\eta_{u_R}-\eta_{\ell_R}\|, \quad 
\|\eta_{d_R}-\eta_{\ell_R}\|, \cr\cr
&& 
\|\eta_{Q_L}-\eta_{\ell_R}\|, \quad 
\|\eta_{\ell,L}-\eta_{u_R}\|, \quad
\|\eta_{\ell,L}-\eta_{d_R}\|, \cr\cr
&& 
\|\eta_{Q_L}-\eta_{L_{\ell,L}}\|  \quad 
{\rm for} \ \ell=e, \ \mu \ . 
\label{nucdec_quark_lepton_distances}
\eeqs
Hence, the full set of lower bounds on fermion separation distances from all of
the inequalities $S^{(Nd)}_r > S^{(Nd)}_{\rm min}$ contributing to 
nucleon decays constitute a set of
coupled quadratic inequalities in the space spanned
by the relevant fermion position vectors. For example, the most stringent lower
bound on a partial lifetime, $(\tau/B)_{p \to e^+ \pi^0}$, yields coupled
quadratic inequalities in the ${\mathbb R}^{5n}$ space spanned by the
vectors $\eta_{Q_L}$, $\eta_{u_R}$, $\eta_{d_R}$,
$\eta_{L_{e,L}}$, and $\eta_{e_R}$, and similarly with nucleon decays involving
$\ell=\mu$. With the inclusion of EW-singlet neutrino fields $\nu_{s,R}$, the
set of separation distances that affect the rates for nucleon decay also
includes
\beq
\|\eta_{q_R}-\eta_{\nu_{s,R}} \|, \quad \|\eta_{Q_L}-\eta_{\nu_{s,R}}\|  
\ {\rm for} \ q=u, \ d \ . 
\label{nucdec_quark_lepton_distances2}
\eeq

The lower bounds on the partial lifetimes for some of the simplest 
proton decays are \cite{abe17} 
\beq
(\tau/B)_{p \to e^+ \pi^0} > 1.6 \times 10^{34} \ {\rm yrs}
\label{tau_bound_p_to_epi0}
\eeq
and
\beq
(\tau/B)_{p \to \mu^+ \pi^0} > 0.77 \times 10^{34} \ {\rm yrs} \ . 
\label{tau_bound_p_to_mupi0}
\eeq
These and the other bounds quoted here are at the 90 \% confidence level.
Other bounds of comparable sensitivity include, e.g., $(\tau/B)_{p \to e^+
  \eta} > 1.0 \times 10^{34}$ yrs, and $(\tau/B)_{p \to \mu^+ \eta} > 0.47
\times 10^{34}$ yrs \cite{abe17d}.  Comparable lower bounds apply for
baryon-number-violating neutron decays, such as $(\tau/B)_{n \to e^+\pi^-} >
0.53 \times 10^{34}$ yr \cite{abe17}, $(\tau/B)_{n \to \mu^+\pi^-} > 0.35
\times 10^{34}$ yr \cite{abe17d}, and $(\tau/B)_{n \to \bar\nu \pi^0} > 1.1
\times 10^{33}$ yr \cite{abe14e} (see also \cite{mcgrew99}).  These bounds can
easily be satisfied by separating the positions of the wavefunction centers of
the quarks and first two generations of leptons \cite{as}.

The calculation of the rate for a nucleon decay to a given final state,
$\Gamma_{N \to f.s.}$, depends on the ultraviolet physics responsible for the
operators ${\cal O}^{(d)}_r$ and their coefficients $\kappa^{(Nd)}_r$ in the
effective Lagrangian.  In particular, it involves the integration of the square
of the matrix element $\langle f.s. | {\cal L}_{eff} | N\rangle$ with respect
to the $n$-body phase space.  Since this ultraviolet physics is not determined
in the context of our low-energy effective Lagrangian approach, it is not
possible to actually perform this integral precisely, but this will not be
necessary for our estimates. Because the
most stringent lower bounds on partial lifetimes of nucleon decays are for
two-body final states, these two-body modes will determine the distance 
constraints, and hence we will only need the two-body phase space factor $R_2$
(see Appendix \ref{phase_space_appendix}). The rate for the decay $N \to f.s.$ 
is 
\begin{widetext}
\beqs
\Gamma_{N \to f.s.} &=& \frac{1}{2m_N} \, \int dR_2 \, |A_{N \to f.s.}|^2  
\cr\cr
                    &=& \frac{1}{2m_N} \, \frac{1}{(M_{BNV})^4} \,
\bigg ( \frac{\mu}{\pi^{1/2} M_{BNV}} \bigg )^{2n} \, \Big | \sum_r 
\bar\kappa^{(Nd)}_r \, e^{-S^{(Nd)}_r} \, 
 \langle f.s. |{\cal O}^{(Nd)}_r | N \rangle \Big |^2 \, R_2 \ , 
\label{gamma_nucdec_gen}
\eeqs
\end{widetext}
where an average over initial spin and sum over final spins is understood. 
As noted above, the dimensionless coefficients $\bar\kappa^{(Nd)}_r$
depend on the UV completion of the extra-dimensional theory and the associated
BSM physics responsible for the baryon number violation, and are not determined
within the framework of our low-energy effective field theory. We take
$\bar\kappa^{(Nd)}_r \simeq O(1)$, and note that it is straightforward to
recalculate bounds on separation distances in the context of a specific UV
completion with different values of the dimensionless coefficients
$\bar\kappa^{(Nd)}_r$. Given these sources of uncertainty, we limit ourselves
to correspondingly rough estimates of lower bounds on fermion separation
distances.  From the most stringent bound on a two-body
proton decay to $\ell^+$ + meson, namely $(\tau/B)_{p \to e^+ \pi^0}$ in 
(\ref{tau_bound_p_to_epi0}), and $(\tau/B)_{p \to \mu^+ \pi^0}$ in 
(\ref{tau_bound_p_to_mupi0}), using estimates of the hadronic matrix elements 
from lattice calculations \cite{saoki_pdec,yaoki_pdec} (and setting 
$\bar\kappa^{(Nd)}_r=1$ as above), we derive the approximate lower bound,
applicable for both of these types of decays, 
\beq
S_r > (S^{(Nd)}_r)_{\rm min} \ , 
\label{srmin2}
\eeq
where
\beqs
(S^{(Nd)}_r)_{\rm min} &=& 48 -\frac{n}{2} \ln \pi 
- 2 \ln \Big ( \frac{M_{BNV}}{100 \ {\rm TeV}} \Big ) \cr\cr
&-& n \, \ln \Big ( \frac{M_{BNV}}{\mu} \Big ) \ . 
\label{srminvalue}
\eeqs
The most direct bound on fermion separation distances arises from the
contribution of the operator $O^{(Nd)}_4$, since the integral $I^{(Nd)}_4$
involves a single fermion separation distance,
$\|\eta_{Q_L}-\eta_{L_{\ell,L}}\|$ for a given lepton generation $\ell=e$ or
$\ell=\mu$. In this case, from the inequality (\ref{srmin2}) with
(\ref{srminvalue}), we obtain the lower bound, for both $e^+$ and $\mu^+$ decay
modes,
\beqs
\| \eta_{Q_L} - \eta_{L_{\ell,L}}\|^2 &>& 62 
- \frac{8}{3}\ln \Big ( \frac{M_{BNV}}{100 \ {\rm TeV}} \Big ) \cr\cr
&-& \frac{8}{3}\ln \Big ( \frac{M_{BNV}}{\mu} \Big ) \ . 
\label{op4_pdec_constraint}
\eeqs
In a model having $n=2$ extra dimensions 
(and value $\mu=3 \times 10^3$ TeV, as given in (\ref{muvalue})), with the
illustrative value $M_{BNV} = 100$ TeV, this is the inequality 
$\| \eta_{Q_L} - \eta_{L_{\ell,L}}\| > 8.4$, while for $M_{BNV}=\mu$, 
this is the inequality $\| \eta_{Q_L} - \eta_{L_{\ell,L}}\| > 7.3$.  
Since $S^{(Nd)}_{\rm min}$ depends only logarithmically on the mass scale
$M_{BNV}$, it follows that the lower bounds on the fermion separation
distances also depend only logarithmically on $M_{BNV}$, i.e., only rather
weakly on this scale.  A very conservative solution to the coupled 
quadratic inequalities would require that each of the relevant distances 
$\|\eta_{f_i}-\eta_{f_j}\|$ in Eq. (\ref{nucdec_quark_lepton_distances}) for
both $\ell=e$ and $\ell=\mu$ would be larger than the square root of the 
right-hand side of Eq. (\ref{srminvalue}):
\begin{widetext}
\beq
\Big \{ \ \|\eta_{u_R}-\eta_{\ell_R}\|, \ 
\|\eta_{d_R}-\eta_{\ell_R}\|, \ 
\|\eta_{Q_L}-\eta_{\ell_R}\|, \ 
\|\eta_{Q_L}-\eta_{L_{\ell,L}}\|, \ 
\|\eta_{L_{\ell,L}}-\eta_{u_R} \|, \ 
\|\eta_{L_{\ell,L}}-\eta_{d_R} \| \ \Big \} 
> [(S^{(Nd)}_r)_{\rm min}]^{1/2} \ . 
\label{nucdec_quark_e_distances}
\eeq
\end{widetext}
That is, this set of inequalities is sufficient, but not necessary, to satisfy
experimental constraints on the model from lower limits on partial lifetimes
for nucleon decays. 

With inclusion of electroweak-singlet $\nu_{s,R}$ fields with small enough 
masses so that they could occur in nucleon decays involving
(anti)neutrinos, an analogous conservative choice would be to impose the same
lower bounds as in Eq. (\ref{nucdec_quark_e_distances}): 
\beqs
&& \Big \{ \ \|\eta_{u_R}-\eta_{\nu_{s,R}} \|, \ 
\|\eta_{d_R}-\eta_{\nu_{s,R}} \|, \ \cr\cr
&&
\|\eta_{Q_L}-\eta_{\nu_{s,R}} \| \ \Big \} > [S^{(Nd)}_{\rm min}]^{1/2}
\label{nucdec_quark_nur_distances}
\eeqs
for all $s$ such that the $\nu_{s,R}$ can occur in nucleon decays. 
We will assume that these inequalities on fermion separation
distances hold in the following.  It is straightforward to use
Eq. (\ref{srbound}) to calculate lower bounds on fermion wavefunction
separation distances with values of $M_{BNV}$ different from the illustrative
value used above.  

The limits on two-body nucleon decays involving (anti)neutrino emission are
somewhat less stringent than the limits on nucleon decays yielding charged
leptons. For example, $(\tau/B)_{p \to \bar\nu \pi^+} > 3.9 \times 10^{32}$ yr
and $(\tau/B)_{n \to \bar\nu \pi^0} > 1.1 \times 10^{33}$ yr
\cite{abe14e}. Hence, they do not add extra information to the constraints that
we have derived on fermion separation distances involving the $L_{\ell,L}$ and
$\ell_R$ fermions with $\ell=e$ or $\ell=\mu$.  However, since a nucleon is
kinematically forbidden from decaying to a real final state containing a $\tau$
lepton, these experimental limits are useful for deriving constraints on
separation distances involving the $L_{\tau,L}$ and $\tau_R$ fermions.  The
relevant operators that would contribute to such decays would be the ${\cal
  O}^{(Nd)}_r$ listed above that contain $L_{\tau,L}$ or $\tau_R$.  The BSM
physics responsible for baryon number violation determines the magnitude of the
corresponding coefficients $\kappa^{(Nd)}_r$.  Since the quark fields in these
four-fermion operators are all of the first generation, a usual expectation
would be that the resultant coefficients for operators in which the lepton
field is of the third generation would be smaller than if the lepton field is
of the first or second generation.  However, to be as conservative as possible,
we consider the possibility of substantial coefficients for such 
four-fermion operators with a third-generation lepton field, namely $\nu_\tau$
(see also \cite{hou2005}). Using the above-mentioned experimental lower bounds
on $(\tau/B)$ for the $p \to \bar\nu \pi^+$ and $n \to \bar\nu \pi^0$ decays in
conjunction with Eqs. (\ref{sdif}) and (\ref{srminvalue}), we obtain the bound
$(S^{(Nd)}_r)_{min,\tau^+} \simeq (S^{(Nd)}_r)_{min}-2$, where
$(S^{(Nd)}_r)_{min}-2$ refers to decay modes such as $p \to e^+ \pi^0$ and $p
\to \mu^+ \pi^0$ and was given in Eq. (\ref{srminvalue}). This can be satisfied
conservatively with the inequality
\begin{widetext}
\beq
\Big \{ \ \|\eta_{q_R}-\eta_{\tau_R}\|, \
\|\eta_{Q_L}-\eta_{\tau_R}\|, \
\|\eta_{q_R}-\eta_{L_{\tau,L}}\|, \ 
\|\eta_{Q_L}-\eta_{L_{\tau,L}}\| \ \Big \} 
> [(S^{(Nd)}_r)_{min,\tau^+}]^{1/2} \quad {\rm for} \ q=u, \ d \ . 
\label{nucdec_quark_tau_distances}
\eeq
\end{widetext}
%

% ====================================================================

\section{$n - \bar n$ Oscillations and Dinucleon Decays to Hadronic Final
  States} 
\label{nnbar_section}

In this section we review the striking finding in Ref. \cite{nnb02}, that in
this extra-dimensional model, even with nucleon decays suppressed well below
experimental limits, $n-\bar n$ oscillations can occur near to their
experimental limits.  Thus, let us consider a general theory in which BSM
physics leads to $n-\bar n$ transitions and let us denote the relevant
low-energy effective Lagrangian, in 4D as ${\cal L}^{(n\bar n)}_{eff}$, and the
transition matrix element $|\delta m| = |\langle \bar n |{\cal L}^{(n\bar
  n)}_{eff} | n \rangle|$.  In (field-free) vacuum, an initial state which is
$|n\rangle$ at time $t=0$ has a nonzero probability to be an $|\bar n\rangle$
state at a later time $t > 0$.  This probability is given by $P(n(t)=\bar n) =
|\langle \bar n|n(t) \rangle|^2 = [\sin^2(t/\tau_{n \bar n})]e^{-t/\tau_n}$,
where $\tau_n$ is the mean life of the neutron. The current direct limit on
$\tau_{n \bar n}$ is from an experiment with a neutron beam from a nuclear
reactor at the Institut Laue-Langevin (ILL) in Grenoble: $\tau_{n \bar n} \ge
0.86 \times 10^8$ sec, i.e., $|\delta m| = 1/\tau_{n \bar n} < 0.77 \times
10^{-29}$ MeV, \cite{ill}.

As noted above, a nonzero $n-\bar n$ transition amplitude $\langle \bar n
|{\cal L}_{eff}|n\rangle$ has the consequence that the resultant physical
eigenstate for the neutron state in matter has a small component of $\bar n$,
i.e., $|n\rangle_{\rm phys.} = \cos\theta_{n\bar n} |n\rangle + \sin\theta_{n
  \bar n} |\bar n\rangle$.  The nonzero $|\bar n\rangle$ component in
$|n\rangle_{\rm phys.}$ leads to annihilation with an adjacent neutron or
proton, and hence to the decays to zero-baryon, multi-meson final states,
consisting dominantly of several pions: $nn \to {\rm pions}$ and $np \to {\rm
  pions}$.  A number of experiments have searched for the resultant matter
instability due to these dinucleon decays and have set lower limits on the
matter instability (m.i.) lifetime, $\tau_{\rm m.i.}$, 
\cite{frejus}-\cite{sk_nnbar}. This is related to
$\tau_{n \bar n}$ by the formula $\tau_{m.i.} = R \, \tau_{n \bar n}^2$, where
$R \sim O(10^2)$ MeV, or equivalently, $R \simeq 10^{23}$ sec$^{-1}$, depending
on the nucleus. The best current limit on matter instability is from the
SuperKamiokande (SK) water Cherenkov experiment \cite{sk_nnbar},
\beq
\tau_{m.i.} > 1.9 \times 10^{32} \ {\rm yr} \ . 
\label{tau_mi}
\eeq
Using the value $R \simeq 0.52
\times 10^{23}$ sec$^{-1}$ for the ${}^{16}$O nuclei in water (see, e.g., 
\cite{nnbar_physrep} and references therein), the SK experiment gives the lower
limit  
\beq
\tau_{n \bar n} > 2.7 \times 10^8 \ {\rm sec} \ , 
\label{tau_nnb}
\eeq
or equivalently, 
\beq
|\delta m | < 2.4 \times 10^{-30} \ {\rm MeV} \ .
\label{delta_m_upper}
\eeq
This lower bound on $\tau_{n \bar n}$ in (\ref{tau_nnb}) from the
SK experiment \cite{sk_nnbar} is comparable to, and
stronger by approximately a factor of 3 than, the direct lower bound on
$\tau_{n \bar n}$ from the ILL experiment \cite{ill}.  The SK experiment
has also searched for specific dinucleon decays and has obtained the
limits \cite{sk_dinucleon_to_pions}  
\beq
\Gamma^{-1}_{np \to \pi^+\pi^0} > 1.70 \times 10^{32} \ {\rm yrs}
\label{tau_np_to_pippi0}
\eeq
and
\beq
\Gamma^{-1}_{nn \to \pi^0\pi^0} > 4.04 \times 10^{32} \ {\rm yrs} \ . 
\label{tau_nn_to_pi0pi0}
\eeq
An improvement in the search for $n - \bar n$ oscillations is anticipated if a
new $n - \bar n$ search with requisite sensitivity could be carried out at the
European Spallation Source (ESS) \cite{nnbar_physrep}.

The effective Lagrangian (in four-dimensional spacetime) that
mediates $n-\bar n$ oscillations is a sum of six-quark operators,
\beq
{\cal L}^{(n \bar n)}_{eff}(x) = \sum_{r=1}^4 c^{(n \bar n)}_r \, 
{\cal O}^{(n \bar n)}_r(x) + h.c. \ . 
\label{leff_nnbar}
\eeq
As with Eqs. (\ref{leff_pd}) and (\ref{leff_higherdim_pd}), there is a
corresponding Lagrangian in the $(4+n)$-dimensional space, 
\beq
{\cal L}^{(n \bar n)}_{eff,4+n}(x,y) = 
\sum_r \kappa^{(n \bar n)}_r O^{(n \bar n)}_r(x,y) + h.c. \ . 
\label{leff_higherdim_nnbar}
\eeq
Since the mass scale characterizing the $|\Delta B|=2$ baryon number violation
is large compared with the electroweak symmetry-breaking scale, these six-quark
operators must be invariant under the Standard-Model gauge symmetry. 
As indicated in Eq. (\ref{leff_nnbar}), there are four 
${\cal O}^{(n \bar n)}_r$ of this type, namely 
\beq
{\cal O}^{(n \bar n)}_1 = (T_s)_{\alpha\beta\gamma\delta\rho\sigma}
[u_R^{\alpha T} C u_R^\beta]
[d_R^{\gamma T} C d_R^\delta]
[d_R^{\rho T} C d_R^\sigma ]
\label{op1_nnbar}
\eeq
\beq
{\cal O}^{(n \bar n)}_2 = (T_s)_{\alpha\beta\gamma\delta\rho\sigma}
[u_R^{\alpha T} C d_R^\beta]
[u_R^{\gamma T} C d_R^\delta]
[d_R^{\rho T} C d_R^\sigma ]
\label{op2_nnbar}
\eeq
\beqs
{\cal O}^{(n \bar n)}_3 &=& 
\epsilon_{ij}(T_a)_{\alpha\beta\gamma\delta\rho\sigma}
[Q_L^{i \alpha T} C Q_L^{j \beta}]
[u_R^{\gamma T} C d_R^\delta]
[d_R^{\rho T} C d_R^\sigma ] \cr\cr
&=& 2(T_a)_{\alpha\beta\gamma\delta\rho\sigma} 
[u_L^{\alpha T} C d_L^{\beta}]
[u_R^{\gamma T} C d_R^\delta]
[d_R^{\rho T} C d_R^\sigma ] \cr\cr
&&
\label{op3_nnbar}
\eeqs
and
\beqs
&& {\cal O}^{(n \bar n)}_4 = \epsilon_{ij}\epsilon_{km}
(T_a)_{\alpha\beta\gamma\delta\rho\sigma} \times \cr\cr
&& \times [Q_L^{i \alpha T} C Q_L^{j \beta}]
[Q_L^{k \gamma T} C Q_L^{m \delta}]
[d_R^{\rho T} C d_R^\sigma ] \cr\cr
&=& 4(T_a)_{\alpha\beta\gamma\delta\rho\sigma}
[u_L^{\alpha T} C d_L^{\beta}]
[u_L^{\gamma T} C d_L^{\delta}]
[d_R^{\rho T} C d_R^\sigma ] \ , \cr\cr
&& 
\label{op4_nnbar}
\eeqs
where, as before, Greek indices $\alpha, \ \beta,...$ are SU(3)$_c$ color
indices; $i,j...$ are weak SU(2)$_L$ indices; and the SU(3)$_c$ color tensors
are 
\beqs
(T_s)_{\alpha \beta \gamma \delta \rho \sigma} &=&
\epsilon_{\rho \alpha \gamma}\epsilon_{\sigma \beta \delta} +
\epsilon_{\sigma \alpha \gamma}\epsilon_{\rho \beta \delta} + \cr\cr
&+&
\epsilon_{\rho \beta \gamma}\epsilon_{\sigma \alpha \delta} +
\epsilon_{\sigma \beta \gamma}\epsilon_{\rho \alpha \delta}
\label{ts}
\eeqs
and
\beq
(T_a)_{\alpha \beta \gamma \delta \rho \sigma} =
\epsilon_{\rho \alpha \beta}\epsilon_{\sigma \gamma \delta} +
\epsilon_{\sigma \alpha \beta}\epsilon_{\rho \gamma \delta} \ . 
\label{ta}
\eeq
(See Appendix \ref{colortensor_appendix} for the symmetry properties of these
tensors.)  

To each of these operators there is a corresponding $V^{(n \bar n)}_r$ 
function, as defined by
Eq. (\ref{uvfactorization}). For example,  
\beqs
&& V^{(n \bar n)}_1 = V^{(n \bar n)}_2 = \cr\cr
&=&  A^6 \exp \Big [ - \Big \{ 
2\|\eta-\eta_{u_R}\|^2 +  4\|\eta-\eta_{d_R}\|^2 \Big \} \Big ] \ , \cr\cr
&& 
\label{chiprod_op4_nnbar}
\eeqs
and so forth for the other two operators.  The resultant integrals
(\ref{integral_r}) over the extra $n$ dimensions comprise three classes.
The integration of the $V^{(n \bar n)}_r$ functions for the operators ${\cal
  O}^{(n \bar n)}_r$ with $r=1,2$ are the same, defining class $C^{(n
  \bar n)}_1$:
\beq
I^{(n \bar n)}_{C_1} = b_6 \, \exp \bigg [ -\frac{4}{3}
\|\eta_{u_R}-\eta_{d_R}\|^2 \bigg ] \ ,  
\label{int_class1_nnbar}
\eeq
where $b_6 = (2 \cdot 3^{-1/2} \, \pi^{-1} \mu ^2)^n$
from the $k=6$ special case of Eq. (\ref{bk}) and 
$I^{(n \bar n)}_{C_k} \equiv I_{C^{(n \bar n)}_k}$. 
The operator ${\cal O}^{(n \bar n)}_3$ yields a second class, 
\begin{widetext}
\beq
I^{(n \bar n)}_{C_2} = b_6 \, \exp \bigg [ -\frac{1}{6}\Big \{ 
2\|\eta_{Q_L}-\eta_{u_R}\|^2 + 
6\|\eta_{Q_L}-\eta_{d_R}\|^2 + 
3\|\eta_{u_R}-\eta_{d_R}\|^2 \Big \} \bigg ] \ . 
\label{int_class2_nucdec}
\eeq
\end{widetext}
Finally, the operator ${\cal O}^{(n \bar n)}_4$ yields the third class, 
\beq
I^{(n \bar n)}_{C_3} = b_6 \, \exp \bigg [ -\frac{4}{3}
\|\eta_{Q_L}-\eta_{d_R}\|^2 \bigg ] \ . 
\label{int_class3_nnbar}
\eeq
From the $k=6$ special cases of Eqs. (\ref{integral_r})-(\ref{crgen}), it
follows that 
\beq
c^{(n \bar n)}_r = \frac{\bar\kappa^{(n \bar n)}_r}{(M_{BNV})^5} \, 
\bigg ( \frac{2\mu^2}{3^{1/2}\pi M_{BNV}^2} \bigg )^n \,  
e^{-S^{(n \bar n)}_r} \ . 
\label{cr_nnbar}
\eeq
where 
\beq
S^{(n \bar n)}_r = \frac{4}{3} \|\eta_{u_R}-\eta_{d_R}\|^2 \quad {\rm for} \
r=1,2, 
\label{srnnb_r12}
\eeq
\beqs
S^{(n \bar n)}_3 &=& \frac{1}{6} \Big \{
2\|\eta_{Q_L}-\eta_{u_R}\|^2 + 
6\|\eta_{Q_L}-\eta_{d_R}\|^2 + \cr\cr
&+& 3\|\eta_{u_R}-\eta_{d_R}\|^2 \Big \} 
\label{srnnb_r3}
\eeqs
and
\beq
S^{(n \bar n)}_4 = \frac{4}{3} \|\eta_{Q_L}-\eta_{d_R}\|^2 \ . 
\label{srnnb_r4}
\eeq
Then
\begin{widetext}
\beq
|\delta m| = \frac{1}{(M_{BNV})^5} \, \Big ( \frac{\mu}{M_{BNV}} \Big )^{2n}
\, \Big (\frac{2}{3^{1/2} \pi} \Big )^n \, 
\Big |\sum_r \bar\kappa^{(n \bar n)}_r \, 
e^{-S^{(n \bar n)}_r} \, \langle n | {\cal O}^{(n \bar n)}_r | n\rangle 
\Big | \ . 
\label{deltam_calc}
\eeq
\end{widetext}
Ref. \cite{nnb02} used, as a specific framework, a model with
$n=2$ and, in addition to the values of $\|\eta_{Q_L}-\eta_{u_R} \|$ and
$\|\eta_{Q_L}-\eta_{d_R} \|$ from (\ref{mf_distance_constraint}), also the
value $\|\eta_{u_R}-\eta_{d_R}\|=7$ from \cite{ms}. It was shown in
\cite{nnb02} that, with this input, the contributions of the ${\cal O}^{(n \bar
  n)}_r$ with $r=1, \ 2, \ 3$ are small compared with the contribution of
${\cal O}^{(n \bar n)}_4$.  Hence, 
$|\delta m| = |c^{(n \bar n)}_4 \, 
\langle \bar n | {\cal O}^{(n \bar n)}_4 | n \rangle |$,
i.e., only the $r=4$ term in Eq. (\ref{deltam_calc}) is non-negligible. 
The sum $S^{(n \bar n)}_4$ is fixed, via Eq. (\ref{mf_distance_constraint}), by
the $d$ quark mass, so, for the given $\mu$ and an input value of $M_{BNV}$
(and with $\kappa^{(n \bar n)}_4 \simeq 1$), the coefficient $c^{(n \bar n)}_4$
is also fixed. The matrix elements $\langle \bar n | {\cal O}^{(n \bar n)}_r |
n \rangle$ have dimensions of $({\rm mass})^6$, and since they are determined
by hadronic physics, one expects on general grounds that they are $\sim
\Lambda_{QCD}^6$, where, as above, $\Lambda_{QCD} \simeq 0.25$ GeV. This is
borne out by quantitative studies \cite{nnb82,nnb84,nnblgt}.  Requiring that
$|\delta m|$ must be less than the experimental upper bound
(\ref{delta_m_upper}) yields a lower bound on $M_{BNV}$ (denoted $M_X$ in
\cite{nnb02}).  With the illustrative value $n=2$, this is
\begin{widetext}
\beq
M_{BNV} > (44 \ {\rm TeV}) \Big ( 
\frac{\tau_{n \bar n}}{2.7 \times 10^8 \ {\rm sec}} \Big )^{1/9} \, 
\Big ( \frac{\mu}{3 \times 10^3 \ {\rm TeV} } \Big )^{4/9} \, 
\bigg ( \frac{|\langle \bar n | {\cal O}^{(n \bar n)}_4 | n\rangle | }
     {\Lambda_{QCD}^6} \bigg )^{1/9}  \ . 
\label{M_nnbar_min}
\eeq
\end{widetext}
Thus, as pointed out in \cite{nnb02}, for values of values of $M_{n \bar n}$ in
the range relevant to our extra-dimensional model, although nucleon 
decays could easily be suppressed well below experimental limits, $n-
\bar n$ oscillations could occur at a level comparable to current limits.

Since the value of the separation distance $\|\eta_{u_R}-\eta_{d_R}\|$ is not
determined by quark masses or mixing (since these arise from bilinear operator
products of $Q_L$ with $u_R$ and $d_R$), it is of interest to inquire what
range of values this distance can have, subject to the condition that $|\delta
m|$ be smaller than the experimental upper limit, (\ref{delta_m_upper}).  With
the input value of $\mu$ given in Eq. (\ref{muvalue}) and for a value of
$M_{BNV} = 50$ TeV, we find the bound $\|\eta_{u_R}-\eta_{d_R}\| \gsim 4.6$.
As noted in Sec. \ref{pdecay_section}, because constraints on fermion
separation distances enter in the sums $S^{(n \bar n)}_r$, the lower bounds on
these distances depend only rather weakly (logarithmically) on $M_{BNV}$.

% ========================================================================

\section{$\Delta L=0$ Dinucleon Decays to Dileptons }
\label{dnd_section} 

The same baryon-number-violating physics that leads to $n-\bar n$ oscillations
and hence also to the dinucleon decays $nn \to {\rm pions}$ and $np \to {\rm
  pions}$ also leads to dinucleon decays to dilepton final states.  These 
decays are of several different types, characterized by 
different $\Delta L$ values: $\Delta L=0$, $\Delta L=-2$, and $\Delta L=2$. 
The $\Delta L=0$ dinucleon decays are on a different footing from the 
$\Delta L= \pm 2$ decays, because a $\Delta L=0$ dinucleon decay can occur via
a combination of a $\Delta B=-1$ $n-\bar n$ transition followed by 
Standard-Model processes, namely the annihilation of the $\bar n$ (i) with
a neighboring $n$ to produce, respectively, a virtual photon or $Z$ which
then creates into a final-state $\ell^+\ell^-$ or $\nu_\ell \bar\nu_\ell$,
or (ii) with a neighboring $p$ to produce a virtual $W^+$, which then creates
the final-state $\ell^+ \bar\nu_\ell$.  

In \cite{dnd} we calculated rough lower bounds on the partial lifetimes for the
above $\Delta L=0$ dinucleon-to-dilepton decays by relating their rates to the
rates for the decays $nn \to \pi^0\pi^0$, $nn \to \pi^+\pi^-$, and $np \to
\pi^+\pi^0$ and using experimental lower bounds on the partial lifetimes of the
latter dinucleon decays.  Our study in \cite{dnd} was a general
phenomenological analysis and did not assume a particular BSM theory such as
the extra-dimension model used in the present work. We obtained the estimated
lower bounds
\beq
(\tau/B)_{nn \to \ell^+\ell^-} \gsim 5 \times 10^{34} \ {\rm yr} \ 
{\rm for} \ \ell=e, \ \mu \ 
\label{tau_limit_nn_to_ellellbar}
\eeq
\beq
(\tau/B)_{nn \to \nu_\ell \bar\nu_\ell} \gsim 2 \times 10^{41} \ {\rm yr}
\quad {\rm for} \ \nu_\ell= \nu_e, \ \nu_\mu, \ \nu_\tau 
\label{tau_limit_nn_to_nunubar}
\eeq
\beq
(\tau/B)_{np \to \ell^+ \nu_\ell}  \gsim 10^{41} \ {\rm yr}
\quad {\rm for} \ \ell=e, \ \mu
\label{tau_limit_np_to_ellnu}
\eeq
and
\beq
(\tau/B)_{np \to \tau^+ \nu_\tau} \gsim 10^{42} \ {\rm yr} \ .
\label{tau_limit_np_to_taunu}
\eeq
These bounds are considerably stronger than the corresponding experimental
bounds from searches for these decays. Experiments use the notational
convention of referring to their limits as limits on $(\tau/B)$ for $nn \to
\pi^0\pi^0$, $n \to \pi^+\pi^-$, and $np \to \pi^+\pi^0$ although their limits
actually apply to the nuclei in their detectors. We follow this convention
here. These experimental bounds are as follows: $(\tau/B)_{nn \to e^+ e^-} >
4.2 \times 10^{33}$ yr and $(\tau/B)_{nn \to \mu^+ \mu^-} > 4.4 \times 10^{33}$
yr from SK \cite{sussman18} (per ${}^{16}O$ nucleus in the water);
$(\tau/B)_{nn \to {\rm inv.}} > 1.4 \times 10^{30}$ yr from KamLAND
\cite{kamland_nn_to_nunubar,kamland_method} (per ${}^{12}$C nucleus in the
liquid scintillator), and $(\tau/B)_{np \to e^+x } > 2.6 \times 10^{32}$ yr,
$(\tau/B)_{np \to \mu^+x } > 2.2 \times 10^{32}$ yr \cite{takhistov15} (per
${}^{16}$O nucleus), where $x$ denotes a neutrino or antineutrino.  Ref.
\cite{bryman} used data from searches for dinucleon decays into multilepton
final states involving $e^+$ and $\mu^+$ plus (anti)neutrinos to obtain the
bound $(\tau/B)_{np \to \tau^+\bar\nu_\tau} > 1 \times 10^{30}$ yrs.  A
dedicated search by SK experiment yielded the bound \cite{takhistov15}
$(\tau/B)_{np \to \tau^+ x} > 2.9 \times 10^{31}$ yr, where, as above, $x$ is a
neutral, weakly interacting fermion, assumed to have a negligibly small
mass. This subsumes the cases in which $x$ is an electroweak-doublet neutrino
or antineutrino of some undetermined flavor, or possibly an electroweak-singlet
(sterile) neutrino. 

% =======================================================================

\section{$\Delta L=-3$ Nucleon Decays to Trileptons }
\label{nuc3_decay_section} 

In this section we consider the $\Delta L=-3$ nucleon decays to trileptons
(\ref{p_to_l2nubar}) and (\ref{n_to_3nubar}).  We use the constraints on
distances derived in Sec. \ref{pdecay_section} to obtain generic expectations
for lower bounds on partial lifetimes for these decays in the extra-dimensional
model. Operators that contribute to the decays (\ref{p_to_l2nubar}) and
(\ref{n_to_3nubar}) are six-fermion operators.  In terms of fermion fields,
the operators that we discuss comprise eight classes, which are listed in Table
\ref{nuc3_class_table}. We denote these with a superscript $(pm3)$, $(nm3)$, or
$(pm3,nm3)$, corresponding to the decays (\ref{p_to_l2nubar}) and
(\ref{n_to_3nubar}) to which the operator contributes, where $pm3$ stands for
``\underline{p}roton decay to tripleptons, with $\Delta L$ equal to
\underline{m}inus \underline{3}'' and similarly for $nm3$. We list these
operators below (with $\ell=e$ or $\mu$), together with the class to which they
belong:
\beqs
{\cal O}^{(pm3)}_1 &=& \epsilon_{\alpha\beta\gamma} 
[u^{\alpha \ T}_R C d^\beta_R]
[u^{\gamma \ T}_R C \ell_R]
[\nu_{s,R}^T C \nu_{s',R}] \ \in C^{(pm3)}_1 \cr\cr
&&
% u^2 d l nu^2  C1 
\label{op1_nuc3}
\eeqs
\beqs
{\cal O}^{(pm3)}_2 &=&  \epsilon_{\alpha\beta\gamma}
[u^{\alpha \ T}_R C d^\beta_R]
[u^{\gamma \ T}_R C \nu_{s,R}]
[\ell_R^T C \nu_{s',R}] \ \in C^{(pm3)}_1 \cr\cr
&&
% u^2 d l nu^2  C1 
\label{op2_nuc3}
\eeqs
\beqs
{\cal O}^{(nm3)}_3 &=& \epsilon_{\alpha\beta\gamma} 
[u^{\alpha \ T}_R C d^\beta_R]
[d^{\gamma \ T}_R C \nu_{s,R}]
[\nu_{s',R}^T C \nu_{s'',R}] \ \in C^{(nm3)}_2 \cr\cr
&&
% u d^2 nu^3 C2 
\label{op3_nuc2}
\eeqs
\beqs
&& {\cal O}^{(pm3)}_4 =  \epsilon_{ij} \epsilon_{\alpha\beta\gamma}
[Q^{i \alpha \ T}_L C Q^{j \beta}_L]
[u^{\gamma \ T}_R C \ell_R]
[\nu_{s,R}^T C \nu_{s',R}] = \cr\cr
&=& 2\epsilon_{\alpha\beta\gamma}
[u^{\alpha \ T}_L C d^{\beta}_L]
[u^{\gamma \ T}_R C \ell_R]
[\nu_{s,R}^T C \nu_{s',R}] \ \in C^{(pm3)}_3 \cr\cr
&&
% Q^2 u l nu^2 C3 
\label{op4_nuc3}
\eeqs
\beqs
&& {\cal O}^{(pm3)}_5 = \epsilon_{ij} \epsilon_{\alpha\beta\gamma}
[Q^{i \alpha \ T}_L C Q^{j \beta}_L]
[u^{\gamma \ T}_R C \nu_{s,R}]
[\ell_R^T C \nu_{s',R}] = \cr\cr
&=& 2\epsilon_{\alpha\beta\gamma}
[u^{\alpha \ T}_L C d^{\beta}_L]
[u^{\gamma \ T}_R C \nu_{s,R}]
[\ell_R^T C \nu_{s',R}] \ \in C^{(pm3)}_3 \cr\cr
&&
\label{op5_nuc3}
\eeqs
% Q^2 u l nu^2  C3
%
%
\beqs
&& {\cal O}^{(nm3)}_6 = \epsilon_{ij} \epsilon_{\alpha\beta\gamma}
[Q^{i \alpha \ T}_L C Q^{j \beta}_L]
[d^{\gamma \ T}_R C \nu_{s,R}]
[\nu_{s',R}^T C \nu_{s'',R}] = \cr\cr
&=& 2\epsilon_{\alpha\beta\gamma}
[u^{\alpha \ T}_L C d^{\beta}_L]
[d^{\gamma \ T}_R C \nu_{s,R}]
[\nu_{s',R}^T C \nu_{s'',R}] \in C^{(nm3)}_4 \cr\cr
&&
\label{op6_nuc3}
\eeqs
% Q^2 d nu^3   C4
%
%
\beqs
&& {\cal O}^{(pm3,nm3)}_7 =  \epsilon_{ij} \epsilon_{\alpha\beta\gamma}
[Q^{i \alpha \ T}_L C L^j_{\ell,L}]
[u^{\beta \ T}_R C d^\gamma_R]
[\nu_{s,R}^T C \nu_{s',R}] = \cr\cr
&=& \epsilon_{\alpha\beta\gamma}
\Big ( [u^{\alpha \ T}_L C \ell_L] - [d^{\alpha \ T}_L C \nu_{\ell,L}] \Big )
[u^{\beta \ T}_R C d^\gamma_R]
[\nu_{s,R}^T C \nu_{s',R}] \cr\cr
&& \in C^{(pm3,nm3)}_5
\label{op7_nuc3}
\eeqs
% Q L u d nu^2  C5 
%
\beqs
&& {\cal O}^{(pm3,nm3)}_8 =  \epsilon_{ij} \epsilon_{\alpha\beta\gamma}
[Q^{i \alpha \ T}_L C L^j_{\ell,L}]
[u^{\beta \ T}_R C \nu_{s,R}]
[d^{\gamma \ T} C \nu_{s',R}] = \cr\cr
&=& \epsilon_{\alpha\beta\gamma}
\Big ( [u^{\alpha \ T}_L C \ell_L] - [d^{\alpha \ T}_L C \nu_{\ell,L}] \Big )
[u^{\beta \ T}_R C \nu_{s,R}]
[d^{\gamma \ T}_R C \nu_{s',R}] \cr\cr
&& \in C^{(pm3,nm3)}_5
\label{op8_nuc3}
\eeqs
% Q L u d nu^2  C5 
%
%
\beqs
&& {\cal O}^{(pm3,nm3)}_9 =  \epsilon_{ij} \epsilon_{\alpha\beta\gamma}
[Q^{i \alpha \ T}_L C L^j_{\ell',L}]
[u^{\beta \ T}_R C \ell_R]
[u^{\gamma \ T}_R C \nu_{s,R}] = \cr\cr
&=& \epsilon_{\alpha\beta\gamma}
\Big ( [u^{\alpha \ T}_L C \ell'_L] - [d^{\alpha \ T}_L C \nu_{\ell',L}] \Big )
[u^{\beta \ T}_R C \ell_R]
[u^{\gamma \ T}_R C \nu_{s,R}] \ . \cr\cr
&& \in C^{(pm3,nm3)}_6
\label{op9_nuc3}
\eeqs
% Q L u^2 l nu  C6
%
%
\beqs
&& {\cal O}^{(pm3,nm3)}_{10} = 
\epsilon_{ij} \epsilon_{km} \epsilon_{\alpha\beta\gamma}
[Q^{i \alpha \ T}_L C Q^{j \beta}_L]
[Q^{k \gamma \ T}_L C L^m_{\ell,L}]
[\nu_{s,R}^T C \nu_{s',R}] = \cr\cr
&=& 2\epsilon_{\alpha\beta\gamma}
[u^{\alpha \ T}_L C d^{\beta}_L]
\Big ( [u^{\gamma \ T}_L C \ell_L] - [d^{\gamma \ T}_L C \nu_{\ell,L}] \Big )
[\nu_{s,R}^T C \nu_{s',R}] \cr\cr
&& \in C^{(pm3,nm3)}_7
\label{op10_nuc3}
\eeqs
% Q^3 L nu^2 C7
%
and
\begin{widetext}
\beqs
{\cal O}^{(pm3,nm3)}_{11} &=&  \epsilon_{ij} \epsilon_{km} 
\epsilon_{\alpha\beta\gamma}
[Q^{i \alpha \ T}_L C L^j_{\ell,L}]
[Q^{k \beta \ T}_L C L^m_{\ell',L}][u^{\gamma \ T}_R C \nu_{s,R}] \cr\cr
&=& \epsilon_{\alpha\beta\gamma}
\Big ( [u^{\alpha \ T}_L C \ell_L] - [d^{\alpha \ T}_L C \nu_{\ell,L}] \Big )
\Big ( [u^{\beta  \ T}_L C \ell'_L] -[d^{\beta  \ T}_L C \nu_{\ell',L}] \Big )
[u^{\gamma \ T}_R C \nu_{s,R}] \ \in C^{(pm3,nm3)}_8 \ . 
\label{op11_nuc3}
\eeqs
% Q^3 L nu^2 C8
%
\end{widetext}
The contributions of the operators are determined by the integrals over the
$n$ extra dimensions, which, in turn, only depend on the class to which a
given operator belongs.  A general remark relevant for these operators and
also operators for other BNV processes is the following: in enumerating
relevant operators contributing to some process, it is sometimes of interest to
demonstrate that they are all linearly independent.  However, for our present
purposes, this is not necessary, since our actual analysis is based on the
classes of operators and their resultant integrals, and these classes are
manifestly independent of each other, since they are comprised of different
fermion fields.  This remark is also relevant for relations involving other 
operators with different Dirac structure. 

Using our general formula (\ref{integral_k_gen}), we calculate the integrals for
these classes.  With the notation $I^{(pm3)}_{C_1} \equiv I_{C^{(pm3)}_1}$, 
we have 
\begin{widetext} 
\beqs
I^{(pm3)}_{C_1} &=& 
b_6 \exp \bigg [ -\frac{1}{6} \Big \{ 
2\| \eta_{u_R} - \eta_{d_R} \|^2 +
2\| \eta_{u_R} - \eta_{\ell_R} \|^2 +
2\| \eta_{u_R} - \eta_{\nu_{s,R}} \|^2 + 
2\| \eta_{u_R} - \eta_{\nu_{s',R}} \|^2 +
 \| \eta_{d_R} - \eta_{\ell_R} \|^2 + \cr\cr
&+&
 \| \eta_{d_R} - \eta_{\nu_{s,R}} \|^2 +
 \| \eta_{d_R} - \eta_{\nu_{s',R}} \|^2 +
 \| \eta_{\ell_R}-\eta_{\nu_{s,R}} \|^2 + 
 \| \eta_{\ell_R}-\eta_{\nu_{s',R}} \|^2 + 
 \| \eta_{\nu_{s,R}}-\eta_{\nu_{s',R}} \|^2 \Big \} \bigg ] \cr\cr
&&
\label{integral_nuc3_class1} 
\eeqs
\beqs
I^{(nm3)}_{C_2} &=& b_6 \exp \bigg [ -\frac{1}{6} \Big \{ 
2\| \eta_{u_R} - \eta_{d_R} \|^2 +
 \| \eta_{u_R} - \eta_{\nu_{s,R}} \|^2 +
 \| \eta_{u_R} - \eta_{\nu_{s',R}} \|^2 +
 \| \eta_{u_R} - \eta_{\nu_{s'',R}} \|^2 +
2\| \eta_{d_R} - \eta_{\nu_{s,R}} \|^2 + \cr\cr
&+&
2\| \eta_{d_R} - \eta_{\nu_{s',R}} \|^2 +
2\| \eta_{d_R} - \eta_{\nu_{s'',R}} \|^2 + 
 \| \eta_{\nu_{s,R}}-\eta_{\nu_{s',R}} \|^2 + 
 \| \eta_{\nu_{s,R}}-\eta_{\nu_{s'',R}} \|^2 + 
 \| \eta_{\nu_{s',R}} - \eta_{\nu_{s'',R}} \|^2 \Big \} \bigg ] \cr\cr
&& 
\label{integral_nuc3_class2} 
\eeqs
\beqs
I^{(pm3)}_{C_3} &=& b_6 \exp \bigg [ -\frac{1}{6} \Big \{ 
2\| \eta_{Q_L} - \eta_{u_R} \|^2 +
2\| \eta_{Q_L} - \eta_{\ell_R} \|^2 +
2\| \eta_{Q_L} - \eta_{\nu_{s,R}} \|^2 +
2\| \eta_{Q_L} - \eta_{\nu_{s',R}} \|^2 +
 \| \eta_{u_R} - \eta_{\ell_R} \|^2 + \cr\cr
&+&
 \| \eta_{u_R} - \eta_{\nu_{s,R}} \|^2 +
 \| \eta_{u_R} - \eta_{\nu_{s',R}} \|^2 + 
 \| \eta_{\ell_R}-\eta_{\nu_{s,R}} \|^2 + 
 \| \eta_{\ell_R}-\eta_{\nu_{s',R}} \|^2 + 
 \| \eta_{\nu_{s,R}} - \eta_{\nu_{s',R}} \|^2 \Big \} \bigg ] \cr\cr
&& 
\label{integral_nuc3_class3} 
\eeqs
\beqs
I^{(nm3)}_{C_4} &=& b_6 \exp \bigg [ -\frac{1}{6} \Big \{ 
2\| \eta_{Q_L} - \eta_{d_R} \|^2 +
2\| \eta_{Q_L} - \eta_{\nu_{s,R}} \|^2 +
2\| \eta_{Q_L} - \eta_{\nu_{s',R}} \|^2 +
2\| \eta_{Q_L} - \eta_{\nu_{s'',R}} \|^2 + 
 \| \eta_{d_R} - \eta_{\nu_{s,R}} \|^2 + \cr\cr
&+&
 \| \eta_{d_R} - \eta_{\nu_{s',R}} \|^2 + 
 \| \eta_{d_R} - \eta_{\nu_{s'',R}} \|^2 + 
 \| \eta_{\nu_{s,R}}-\eta_{\nu_{s',R}} \|^2 + 
 \| \eta_{\nu_{s,R}}-\eta_{\nu_{s'',R}} \|^2 + 
 \| \eta_{\nu_{s',R}} - \eta_{\nu_{s'',R}} \|^2 \Big \} \bigg ] \ . \cr\cr
&& 
\label{integral_nuc3_class4} 
\eeqs
\beqs
I^{(pm3,nm3)}_{C_5} &=& b_6 \exp \bigg [ -\frac{1}{6} \Big \{ 
 \| \eta_{Q_L} - \eta_{L_{\ell,L}} \|^2 +
 \| \eta_{Q_L} - \eta_{u_R} \|^2 +
 \| \eta_{Q_L} - \eta_{d_R} \|^2 +
 \| \eta_{Q_L} - \eta_{\nu_{s,R}} \|^2 +
 \| \eta_{Q_L } - \eta_{\nu_{s',R}} \|^2 + \cr\cr
&+&
 \| \eta_{L_{\ell,L}} - \eta_{u_R} \|^2 +
 \| \eta_{L_{\ell,L}} - \eta_{d_R} \|^2 +
 \| \eta_{L_{\ell,L}} - \eta_{\nu_{s,R}} \|^2 +
 \| \eta_{L_{\ell,L}} - \eta_{\nu_{s',R}} \|^2 +
 \| \eta_{u_R} - \eta_{d_R} \|^2 + \cr\cr
&+&
 \| \eta_{u_R} - \eta_{\nu_{s,R}} \|^2 + 
 \| \eta_{u_R} - \eta_{\nu_{s',R}} \|^2 + 
 \| \eta_{d_R} - \eta_{\nu_{s,R}} \|^2 + 
 \| \eta_{d_R} - \eta_{\nu_{s',R}} \|^2 + 
 \| \eta_{\nu_{s,R}} - \eta_{\nu_{s',R}} \|^2 \Big \} \bigg ] \cr\cr
&& 
\label{integral_nuc3_class5} 
\eeqs
\beqs
I^{(pm3,nm3)}_{C_6} &=& b_6 \exp \bigg [ -\frac{1}{6} \Big \{ 
 \| \eta_{Q_L} - \eta_{L_{\ell,L}} \|^2 +
2\| \eta_{Q_L} - \eta_{u_R} \|^2 +
 \| \eta_{Q_L} - \eta_{\ell_R} \|^2 +
 \| \eta_{Q_L} - \eta_{\nu_{s,R}} \|^2 +
2\| \eta_{L_{\ell,R}} - \eta_{u_R} \|^2 + \cr\cr
&+&
 \| \eta_{L_{\ell,L}} - \eta_{\ell_R} \|^2 +
 \| \eta_{L_{\ell,L}} - \eta_{\nu_{s,R}} \|^2 +
2\| \eta_{u_R} - \eta_{\ell_R} \|^2 +
2\| \eta_{u_R} - \eta_{\nu_{s,R}} \|^2 + 
 \| \eta_{\ell_R} - \eta_{\nu_{s,R}} \|^2  \Big \} \bigg ] \cr\cr
&& 
\label{integral_nuc3_class6} 
\eeqs
\beqs
I^{(pm3,nm3)}_{C_7} &=& b_6 \exp \bigg [ -\frac{1}{6} \Big \{ 
3\| \eta_{Q_L} - \eta_{L_{\ell,L}} \|^2 +
3\| \eta_{Q_L} - \eta_{\nu_{s,R}} \|^2 +
3\| \eta_{Q_L} - \eta_{\nu_{s',R}} \|^2 +
 \| \eta_{L_{\ell,L}} - \eta_{\nu_{s,R}} \|^2 + \cr\cr
&+&
 \| \eta_{L_{\ell,L}} - \eta_{\nu_{s',R}} \|^2 +
 \| \eta_{\nu_{s,R}} - \eta_{\nu_{s',R}} \|^2  \Big \} \bigg ] 
\label{integral_nuc3_class7} 
\eeqs
and
\beqs
I^{(pm3,nm3)}_{C_8} &=& b_6 \exp \bigg [ -\frac{1}{6} \Big \{ 
4\| \eta_{Q_L} - \eta_{L_{\ell,L}} \|^2 +
2\| \eta_{Q_L} - \eta_{u_R} \|^2 +
2\| \eta_{Q_L} - \eta_{\nu_{s,R}} \|^2 +
2\| \eta_{L_{\ell,L}} - \eta_{u_R} \|^2 + \cr\cr
&+& 
2\| \eta_{L_{\ell,L}} - \eta_{\nu_{s,R}} \|^2 +
 \| \eta_{u_R} - \eta_{\nu_{s,R}} \|^2  \Big \} \bigg ] \ . 
\label{integral_nuc3_class8} 
\eeqs
\end{widetext} 

Using these calculations and typical values of fermion separation distances
obeying the constraints from nucleon decays discussed in Sec.
\ref{pdecay_section}, we find that these $\Delta L=-3$ nucleon decays are
strongly suppressed relative to nucleon decays mediated by four-fermion
operators.  Making reference to the comparison of rates in
Eq. (\ref{ratio_k64}) and the illustrative numerical example in
Eq. (\ref{log_ratio_k64_example}), we find that the difference $\langle
S_{(6)}\rangle - \langle S_{(4)}\rangle$ is positive, adding to the suppression
from the prefactor.  The basic reason that the $\Delta L=-3$ decays
to trilepton final states are strongly suppressed in this model, while $n-\bar
n$ oscillations can occur at levels comparable to current limits is BNV nucleon
decays can be suppressed by making the separation between quark and lepton
wavefunction centers sufficiently large. This does not suppress $n-\bar n$
oscillations, but considerably suppresses these $\Delta L=-3$ decays, since
they involve outgoing (anti)leptons.  This reason also explains the suppression
that we will find for the various types of BNV nucleon and dinucleon decays in
the following sections.

Thus, we find that the resultant expected predictions for partial lifetimes for
these $\Delta L=-3$ nucleon decays are compatible with existing experimental
limits. These limits include $(\tau/B)_{p \to e^+ x x } > 0.58 \times 10^{30}$
yr \cite{takhistov14}, $(\tau/B)_{p \to \mu^+ x x } > 0.58 \times 10^{30}$ yr
\cite{takhistov14}, and $(\tau/B)_{n \to xxx} > 0.58 \times 10^{30}$ yr
\cite{araki06}, where here $x$ denotes an unobserved neutral, weakly
interacting fermion with negligibly small mass that does not decay in the
detector.  Thus, for example, the lower limit on $(\tau/B)$ for the decay $p
\to \ell^+ x x$ applies to all of the decays $p \to \ell^+ \bar\nu \bar\nu'$
(with $\Delta L=-3$), $p \to \ell^+ \nu \bar\nu'$ (with $\Delta L=-1$), and $p
\to \ell^+ \nu \nu'$ (with $\Delta L=1$) for $\ell^+=e^+$ or $\mu^+$, and
similarly, the lower bound on $n \to xxx$ applies to all neutron decays to
combinations of (anti)neutrinos with $\Delta L$ ranging from $\Delta L=-3$ to
$\Delta L=+3$. Further searches for these and other types of nucleon decays are
worthwhile (e.g., \cite{babu_mohapatra,kobach,heeck_takhistov}).  In addition
to continued data-taking at Superkamiokande, future searches for nucleon decays
are planned at HyperKamiokande \cite{hk} and in the liquid argon detector in
DUNE (Deep Underground Neutrino Experiment) \cite{dune_physics}.

% =====================================================================

\section{$\Delta L=1$ Nucleon Decays to Trileptons} 
\label{nuc3b_decay_section}

Here we study the $\Delta L=1$ nucleon decays to trilepton final states
(\ref{p_to_l2nu}) and (\ref{n_to_nu2nubar}). These decays are mediated by
six-fermion operators, as was the case with the $\Delta L=-3$ nucleon decays to
trilepton final states analyzed in Sec. \ref{nuc3_decay_section}.  Our
procedure for analyzing these decays is analogous to the procedure we used in
Sec. \ref{nuc3_decay_section}. Indeed, there is a 1-1 correspondence between
the operators here and a subset of the operators in that section, namely ${\cal
  O}^{(Nm3)}_r$ with $r=1, \ 3, \ 4, \ 6, \ 7$, obtained by the replacement of
an EW-singlet neutrino bilinear by one with each $\nu_{s,R}$ field replaced by
$(\nu_{s,R})^c \equiv (\nu^c)_{s,L}$ (the charge conjugation reverses the
chirality), i.e., by replacing $[\nu_{s,R}^T C \bar\nu_{s',R}]$ by $[\nu^{c \
  T}_{s,L} C \nu^c_{s',L}]$. We denote these with a superscript $(p1)$, $(n1)$,
or $(p1,n1)$, corresponding to the decays (\ref{p_to_l2nu}) and
(\ref{n_to_nu2nubar}) to which the operator contributes, where $p1$ stands for
``\underline{p}roton decay to tripleptons, with $\Delta L$ equal to
\underline{1}'' and similarly for $n1$. The charge conjugation leaves the
position of the fermion unchanged, so $\eta_{\nu_s,R} = \eta_{\nu^c_{s,L}}$.
Consequently, the five classes to which the operators for the $\Delta L=1$
nucleon decays to trileptons belong are in 1-1 correspondence with five of the
seven classes to which the operators for the $\Delta L=-3$ nucleon decays to
trileptons belong, and the corresponding integrals are equal:
\begin{widetext}
\beq
C^{(p3)}_1 \leftrightarrow C^{(p1)}_1, \quad 
C^{(n3)}_2 \leftrightarrow C^{(n1)}_2, \quad 
C^{(p3)}_3 \leftrightarrow C^{(p1)}_3, \quad 
C^{(n3)}_4 \leftrightarrow C^{(n1)}_4, \quad 
C^{(p3,n3)}_5 \leftrightarrow C^{(p1,n1)}_1
\label{class_equivalence_nucleon_to_trileptons}
\eeq
where here the symbol $\leftrightarrow$ means replacement of a $\nu \nu$
bilinear by a $\nu^c \nu^c$ bilinear. The integrals satisfy the equalities
\beq
I^{(p1)}_{C_1} = I^{(pm3)}_{C_1}, \quad 
I^{(n1)}_{C_2} = I^{(nm3)}_{C_2}, \quad 
I^{(p1)}_{C_3} = I^{(pm3)}_{C_3}, \quad 
I^{(n1)}_{C_4} = I^{(nm3)}_{C_4}, \quad 
I^{(p1,n1)}_{C_5} = I^{(pm3,nm3)}_{C_5}
\label{integral_equivalence_nucleon_to_trileptons} 
\eeq
\end{widetext}
Operators mediating these $\Delta L=1$ dinucleon decays to trileptons are 
%, 
\beqs
{\cal O}^{(p1)}_1 &=& \epsilon_{\alpha\beta\gamma} 
[u^{\alpha \ T}_R C d^\beta_R]
[u^{\gamma \ T}_R C \ell_R]
[\nu^{c \ T}_{s,L} C \nu^c_{s',R}] \ \in C^{(p1)}_1 \cr\cr
&&
% u^2 d l nusbar^2  C1 
\label{op1_nuc3b}
\eeqs
\beqs
{\cal O}^{(n1)}_2 &=& \epsilon_{\alpha\beta\gamma} 
[u^{\alpha \ T}_R C d^\beta_R]
[d^{\gamma \ T}_R C \nu_{s,R}]
[\nu^{c \ T}_{s',L} C \nu^c_{s'',L}] \ \in C^{(n1)}_2 \cr\cr
&&
% u d^2 nu nusbar^2 C2 
\label{op3_nuc3b}
\eeqs
\beqs
&& {\cal O}^{(p1)}_3 =  \epsilon_{ij} \epsilon_{\alpha\beta\gamma}
[Q^{i \alpha \ T}_L C Q^{j \beta}_L]
[u^{\gamma \ T}_R C \ell_R]
[\nu^{c \ T}_{s,L} C \nu^c_{s',L}] = \cr\cr
&=& 2\epsilon_{\alpha\beta\gamma}
[u^{\alpha \ T}_L C d^{\beta}_L]
[u^{\gamma \ T}_R C \ell_R]
[\nu^{c \ T}_{s,L} C \nu^c_{s',L}] \ \in C^{(p1)}_3 \cr\cr
&&
% Q^2 u l nusbar^2 C3 
\label{op4_nuc3b}
\eeqs
\beqs
&& {\cal O}^{(n1)}_4 = \epsilon_{ij} \epsilon_{\alpha\beta\gamma}
[Q^{i \alpha \ T}_L C Q^{j \beta}_L]
[d^{\gamma \ T}_R C \nu_{s,R}]
[\nu^{c \ T}_{s',L} C \nu^c_{s'',L}] = \cr\cr
&=& 2\epsilon_{\alpha\beta\gamma}
[u^{\alpha \ T}_L C d^{\beta}_L]
[d^{\gamma \ T}_R C \nu_{s,R}]
[\nu^{c \ T}_{s',L} C \nu^c_{s'',L}] \ \in C^{(n1)}_4 \cr\cr
&&
\label{op6_nuc3b}
\eeqs
% Q^2 d nu nusbar^2   C4
%
and
\beqs
&& {\cal O}^{(p1,n1)}_5 =  \epsilon_{ij} \epsilon_{\alpha\beta\gamma}
[Q^{i \alpha \ T}_L C L^j_{\ell,L}]
[u^{\beta \ T}_R C d^\gamma_R]
[\nu^{c \ T}_{s,L} C \nu^c_{s',L}] = \cr\cr
&=& \epsilon_{\alpha\beta\gamma}
\Big ( [u^{\alpha \ T}_L C \ell_L] - [d^{\alpha \ T}_L C \nu_{\ell,L}] \Big )
[u^{\beta \ T}_R C d^\gamma_R][\nu^{c \ T}_{s,L} C \nu^c_{s',L}] \cr\cr
&& \in C^{(p1,n1)}_5 \ . 
\label{op7_nuc3b}
\eeqs
% Q L u d nusbar^2  C5 
%
Owing to the equalities (\ref{integral_equivalence_nucleon_to_trileptons}), our
conclusions concerning upper bounds on the rates for these $\Delta L=1$ nucleon
decays to trilepton final states are the same as for the $\Delta L=-3$ nucleon
decays to trileptons.

% ======================================================================

\section{$\Delta L=-2$ Dinucleon Decays to Dileptons: General Operator 
Analysis }
\label{dinucleon_to_dilepton_section}

In this section we carry out a general operator analysis of the $\Delta L=-2$
dinucleon decays to dileptons (\ref{pp_to_ll})-(\ref{nn_to_2nubar}). In later
sections, we shall use our results to obtain approximate estimates of expected
rates for these decays in the extra-dimensional model.  As is obvious from the
selection rule $\Delta L=-2$ for these decays, they arise differently from the
$\Delta B=-2$, $\Delta L=0$ dinucleon-to-dilepton decays for which we set
bounds in \cite{dnd}. The process by which the $\Delta B=-2$, $\Delta L=0$
dinucleon-to-dilepton decays occur involves a local six-fermion operator that
mediates the $n-\bar n$ transition, in conjunction with $n \bar n$ annihilation
leading to a virtual $\gamma$, $Z$, or $\bar n p$ annihilation leading to a
virtual $W^+$.  The virtual $\gamma$, $Z$, or $W+$ then produce the final-state
lepton-antilepton pairs, namely $\ell^+ \ell^-$, $\nu_\ell \bar\nu_\ell$, and
$\ell^+ \nu_\ell$, respectively.  Although the amplitudes involve eight
external fermion lines, the lepton-antilepton operator product is bilocal with
respect to the six-quark operator product (separated by a Euclidean distance
$\sim 1/{\rm fm}$ for the $\gamma$, $\sim 1/m_Z$ and $\sim 1/m_W$ for the
processes with a virtual $Z$ and $W^+$, respectively; i.e., these $\Delta
B=-2$, $\Delta L=0$ amplitudes do not dominantly involve local eight-fermion
operator products.

Proceeding with our analysis, we first discuss the general structure of an 
effective Lagrangian for the $\Delta B=-2$, $\Delta L=-2$ dinucleon-to-dilepton
decays.  For labelling purposes, we shall introduce the superscript $NN'$,
which takes on the respective values $(NN')=(pp)$ for $pp \to \ell^+ \ell'^+$
decays, $(NN')=(np)$ for $np \to \ell^+ \bar\nu$ decays, and $(NN')=(nn)$ for
$nn \to \bar\nu \bar\nu'$ decays, with the dilepton final state kept implicit
in the notation.  This effective Lagrangian has the form 
\beqs
{\cal L}^{(NN')}_{eff}(x) &=& 
\sum_r c_r^{(NN')} {\cal O}^{(NN')}_r(x) + h.c. \cr\cr
 &=& \sum_r \kappa^{(NN')}_r \, \int d^n y \, O^{(NN')}_r(x,y) + h.c. 
\cr\cr
 &=& \sum_r \kappa^{(NN')}_r U^{(NN')}_r(x) \, 
\int d^n y \, V^{(NN')}_r(y) + h.c. \cr\cr
 &=& \sum_r \kappa^{(NN')}_r \, I^{(NN')}_r \, 
U^{(NN')}_r(x) + h.c. \ , 
\label{leff_dinucleon_to_dilepton}
\eeqs
where, in accord with the general notation (\ref{integral_r}), 
\beq
I^{(NN')}_r = \int d^n y \, V^{(NN')}_r(y) \ . 
\label{integral_dinucleon_to_dilepton}
\eeq
Various sets of operators $O^{(NN)}_r$ yield the same
integrals $I^{(NN)}_r$, so they can be organized into certain 
classes, as we will discuss below. 

By the same logic as for the four-fermion operators contributing to individual
nucleon decays and the six-quark operators contributing to $n-\bar n$
oscillations and dinucleon decays to mesonic final states, since existing
limits imply that the mass scale characterizing the physics responsible for
these dinucleon-to-dilepton decays must be large compared with the electroweak
symmetry-breaking scale $v$, it follows that the eight-fermion operators
$O^{(NN')}_r(x,y)$ must be singlets under the Standard-Model gauge group,
$G_{SM}$.

Six of the eight fermions in these operators are quark fields. 
The color indices of the six quark fields, denoted as $\alpha,
\ \beta, \ \gamma, \ \delta, \ \rho, \ \sigma$, are coupled together to make an
SU(3)$_c$ singlet. This can be done in any of three ways, corresponding to the
color tensors $(T_s)_{\alpha \beta \gamma \delta \rho \sigma}$ in
Eq. (\ref{ts}), $(T_a)_{\alpha \beta \gamma \delta \rho \sigma}$ in
Eq. (\ref{ta}), and 
\beq
(T_{a3})_{\alpha \beta \gamma \delta \rho \sigma} =
\epsilon_{\rho \alpha \beta}\epsilon_{\sigma \gamma \delta} -
\epsilon_{\sigma \alpha \beta}\epsilon_{\rho \gamma \delta} \ . 
\label{ta3}
\eeq
Some properties of these tensors are reviewed in Appendix
\ref{colortensor_appendix}.  As discussed in \cite{nnb84}, there are also color
tensors related to these by redefinition of indices, such as $T_{a2'(saa)}$ and
$T_{a2'(asa)}$ in Eqs. (3.4) and (3.5) of \cite{nnb84}, but these will not be
needed here.

The eight-fermion operators can be classified according to how many of the
eight fermions are SU(2)$_L$ nonsinglets; the possibilities are 0, 2, 4, 6, and
8. For operators containing a nonzero number (2, 4, 6, or 8)
fermions in SU(2)$_L$ nonsinglets, there are various ways to contract the
SU(2)$_L$ weak isospin indices.  One way is to contract each pair of 
weak isospin-1/2 indices antisymmetrically to make singlets, using 
the $\epsilon_{ij}$ tensor for two SU(2)$_L$ indices, and so forth for other
SU(2)$_L$ indices. Alternatively, one can 
combine pairs of weak isospin-1/2 fields symmetrically to make adjoint
(i.e., weak isospin 1) representations of SU(2)$_L$ and then contract these
to obtain an SU(2)$_L$ singlet.  For example, starting with four weak isospin
1/2 representations with SU(2)$_L$ indices $(i,j), \ (k,m)$, the $(i,j)$ and
$(k,m)$ indices can each be combined symmetrically, and then
the resulting two isovectors can be contracted to make an SU(2)$_L$
singlet. This is done with the SU(2)$_L$ tensor
\beq
(I_{ss})_{ijkm} \equiv 
(\epsilon_{ik}\epsilon_{jm} + \epsilon_{im}\epsilon_{jk}) \ .
\label{iss}
\eeq
For operators with six fermions in SU(2)$_L$ doublets, another relevant
SU(2)$_L$ tensor involves symmetric combinations of two pairs of isospin-1/2
representations combined with an antisymmetric combination of the third pair of
isospin-1/2 representations, via the tensor
\beq
(I_{ssa})_{ijkmnp} \equiv (
\epsilon_{ik}\epsilon_{jm}+
\epsilon_{im}\epsilon_{jk} ) \epsilon_{np} \ , 
\label{issa}
\eeq
where the subscript $(ssa)$ refers to this symmetric-symmetric-antisymmetric
structure of SU(2)$_L$ contractions.  Finally, one can also use a set of
SU(2)$_L$ contractions in which all pairs of isospin-1/2 representations are
combined symmetrically.  The SU(2)$_L$ tensor that does this is
\beqs
I_{sss} &=& 
\epsilon_{ik}\Big (\epsilon_{jn}\epsilon_{mp}+\epsilon_{mn}\epsilon_{jp}\Big )
+ 
\epsilon_{im}\Big (\epsilon_{jn}\epsilon_{kp}+\epsilon_{kn}\epsilon_{jm}\Big )
\cr\cr
&+& 
\epsilon_{jk}\Big (\epsilon_{in}\epsilon_{mp}+\epsilon_{mn}\epsilon_{ip}\Big )
+
\epsilon_{jm}\Big (\epsilon_{in}\epsilon_{kp}+\epsilon_{kn}\epsilon_{ip}\Big )
\ , 
\cr\cr
&& 
\label{isss}
\eeqs
where the $(sss)$ subscript refers to the threefold symmetric set of
contractions. 

Since there is a 1-1 correspondence between an operator ${\cal O}^{(NN')}_r$ in
${\cal L}^{(NN')}_{eff}$ and an operator $O^{(NN')}_r$ in ${\cal
  L}^{(NN')}_{eff,4+n}$, one can use either of these for a structural
analysis; we will use the ${\cal O}^{(NN')}_r$.  We 
will determine a general set of classes of operators that yield the same
integrals $I^{(NN')}_r$, as defined in Eq.
(\ref{integral_dinucleon_to_dilepton}). A 
given class typically contains several different individual operators.
However, since it is the integrals $I^{(NN')}_r$ that control the contribution
to the amplitude, the natural organization for our analysis is in terms of
these classes, rather than the individual operators.  

We proceed with the general structural analysis of the $\Delta L=-2$
dinucleon-to-dilepton decays. The eight fermions that
comprise a given operator ${\cal O}^{(NN')}_r$ are comprised of six quarks and
  two leptons, namely $uud,uud,\ell^+,\ell'^+$, $uud,ddu,\ell^+ \bar\nu$, and
  $ddu,ddu,\bar\nu,\bar\nu'$ for the decays (\ref{pp_to_ll}),
  (\ref{np_to_lnubar}), and (\ref{nn_to_2nubar}).  As discussed above, the
  quarks can be chosen from the SU(2)$_L$-doublet $Q_L$ or the
  SU(2)$_L$-singlets $u_R$ and $d_R$, and the leptons can be chosen from the
  SU(2)$_L$-doublets $L_{\ell,L}$, $L_{\ell',L}$ and the SU(2)$_L$-singlets
  $\ell_R$, $\ell'_R$, and $\nu_{s,R}$. We can abstractly represent a generic
  eight-fermion operator product ${\cal O}^{(NN')}_r$ as
\beq
{\cal O}^{(NN')} = Q_L^{n_Q} \, L_L^{n_L} \, u_R^{n_u} \, d_R^{n_d} \, 
\ell_R^{n_\ell} \, \nu_{s,R}^{n_{\nu_s}} \ ,   
\label{ogeneral}
\eeq
where we have suppressed the arguments $y$ and $\eta$ in the fermion fields,
have suppressed the difference between lepton fields with and without primes,
and have left the chiralities of the fermions implicit in the exponents. 
The fact that the operator involves eight fermions is the condition
\beq
n_Q + n_L + n_u + n_d + n_\ell + n_{\nu_s} = 8 \ . 
\label{8fermions}
\eeq
The condition that the initial state is a dinucleon is that
\beq
n_Q+n_u+n_d = 2N_c = 6 \ , 
\label{nq_condition}
\eeq
where $N_c=3$ is the number of colors.  With the color contractions discussed
above, this condition is sufficient for the operator be an SU(3)$_c$ singlet.
The condition that the final state has $L=-2$, i.e., is comprised of two
antileptons, is that
\beq
n_L+n_{\ell}+n_{\nu_s}=2 \ .
\label{nl_condition}
\eeq
Note that only two of the three equations (\ref{8fermions}), 
(\ref{nq_condition}), and (\ref{nl_condition}) are linearly independent. 
The requirement that ${\cal O}$ must be invariant under the SM gauge group
implies that it must have zero weak hypercharge and that it must be a singlet
under SU(2)$_L$.  The condition that it must have weak hypercharge $Y=0$ 
is that $\sum_f n_f Y_f=0$, or, explicitly,
\beqs
&& n_Q \Big ( \frac{1}{3} \Big ) + n_L (-1) + n_u \Big ( \frac{4}{3} \Big ) 
+ n_d \Big (-\frac{2}{3} \Big ) + n_\ell (-2) =0 \ . \cr\cr
&& 
\label{yzero_condition}
\eeqs
The condition that the operator must be an SU(2)$_L$ singlet requires that the
number of SU(2)$_L$ doublets must be even:
\beq
n_Q + n_L = (0, \ 2, \ 4,\ 6, \ {\rm or} \ 8) \ . 
\label{su2_invariant_condition}
\eeq
Eqs. (\ref{8fermions}-(\ref{su2_invariant_condition}) comprise five linear
equations, of which four are linearly independent, in the six (non-negative,
integer) unknown numbers, $n_Q$, $n_L$, $n_u$, $n_d$, $n_\ell$, and
$n_{\nu_s}$, with the constraint that each number must lie in the range
$[0,8]$.  The solutions to these equations with the given constraint determines
the general structures of the operators for dinucleon-to-dilepton decays with
$\Delta L=-2$. We have obtained these solutions, which we list in Table
\ref{operator_class_table}.  The abbreviations used for the fermion fields are
$Q = Q_L$, $L=L_L$, $u=u_R$, $d=d_R$, $\ell=\ell_R$, and $\nu = \nu_{s,R}$. The
first column lists the class number; the second column lists the number of
SU(2)$_L$ doublets, denoted $N_d$; and the third column lists the general
structure.  Primes distinguishing different lepton fields are suppressed in the
notation. In checking candidate solutions of Eqs.
(\ref{8fermions}-(\ref{su2_invariant_condition}), it is necessary to verify
that they do not vanish identically because of combined SU(3)$_c$ and SU(2)$_L$
tensors.  We find that one class with $N_d=6$, of the abstract form
$Q^6\ell\nu$, contains no nonvanishing operators of our type.  We denote a
given class symbolically as $C^{(NN')}_k$. These contribute as follows:
\beqs
pp \to \ell^+ \ell'^+: \ && C^{(NN')}_k, \ k=1, \ 4, \ 7, \ 10, \ 13, 
\ 15, \ 16, \cr\cr 
&& 17, \ 19
\label{pp_to_ll_classes}
\eeqs
\beqs
np \to \ell^+ \bar\nu: \ &&  C^{(NN')}_k, \ 
k=2, \ 5, \ 7, \ 8, \ 9, \ 11, \ 13, \cr\cr
&& 14, \ 15, \ 16, \ 17, \ 18, \ 19 
\label{np_to_lnubar_classes}
\eeqs
and
\beqs
nn \to \bar\nu \bar\nu': 
\  && C^{(NN')}_k, \ k=3, \ 6, \ 8, \ 12, \ 14, \ 15, \ 16, \cr\cr
&& \ 18, \ 19 \ .
\label{nn_to_2nubar_classes}
\eeqs
As is evident in these lists, some classes of operators only contribute to one
type of $\Delta L=-2$ dinucleon-to-dilepton decay, while others contribute to 
two or three of these decays. We will sometimes indicate this explicitly,
writing, for example, $C^{(NN')}_1 = C^{(pp)}_1$, $C^{(NN')}_2=C^{(np)}_2$,
$C^{(NN')}_3=C^{(nn)}_3$, $C^{(NN')}_7 = C^{(pp,np)}_7$, 
$C^{(NN')}_8=C^{(np,nn)}_8$, and $C^{(NN')}_{15}=C^{(pp,np,nn)}_{15}$, 
where abbreviations for superscripts are $pp$ for the decays $pp \to
\ell^+ \ell'^+$, $np$ for $np \to \ell^+ \bar\nu$, and $nn$ for $nn \to \bar\nu
\bar\nu'$. For brevity, we will also sometimes suppress the superscript $(NN')$ 
on $C^{(NN')}_k$, writing simply $C_k$, as in the notation 
$I^{(NN')}_{C_k} \equiv I_{C^{(NN')}_k}$.

The integrand function of a class of operators $C^{(NN')}_k$ in this table with 
a given set of exponents $(n_Q,n_L,n_u,n_d,n_\ell,n_\nu)$ is of the form 
\beq
V^{(NN')}_k(y) = A^8 \, \exp \Big [ -\sum_{\{ f \}} n_f 
\|\eta-\eta_f \|^2 \Big ] \ . 
\label{vdinucleon_to_dilepton}
\eeq
The integral of $V^{(NN')}_k(y)$ over the extra spatial coordinates is 
\beq
I^{(NN')}_{C_k}= \int d^n y \, V^{(NN')}_k(y) \ . 
\label{integral_dinucleon_class}
\eeq
This gives 
\beq
I^{(NN')}_{C_k} = 
b_8 \, \exp \bigg [ -\frac{1}{8} \sum_{f,f'; \ f \ne f',
  ord.} n_f n_{f'} \| \eta_f - \eta_{f'} \|^2 \bigg ] \ , 
\label{integral_k_gen}
\eeq
where the sum is over all of the types of fermion fields in the operator
product, in an ordered manner, as indicated in Eq. (\ref{intform}). The
prefactor $b_8 = (2^{1/2}\pi^{-3/2}\mu^3)^n$, from Eq. (\ref{bk}). 
As noted in connection with Eq. (\ref{intform}), for an operator 
$O^{(NN')}_k$ containing $N_f$ different types of fermion fields, the 
integral $I^{(NN')}_{C_k}$ depends on ${N_f \choose 2}$ different separation 
distances $\|\eta_f - \eta_{f'}\|$. 

As the $k=8$ special case of Eq. (\ref{crgen}), the coefficient $c^{(NN')}_r$ 
can be expressed as 
\beqs
c^{(NN')}_r &=& \kappa^{(NN')}_r \, I^{(NN')}_r = 
\frac{\bar\kappa^{(NN')}_r}{(M_{BNV})^{8+3n}} \, 
b_8 \, e^{-S^{(NN')}_r} \cr\cr
&=& \frac{\bar\kappa^{(NN')}_r}{M_{BNV}^8} \, 
\Big ( \frac{2^{1/2} \mu^3}{\pi^{3/2} M_{BNV}^3} \Big )^n \, 
e^{-S^{(NN')}_r} \ , 
\label{cr_dinucleon_to_dilepton}
\eeqs
Then the decay rate for one of the three dinucleon-to-dilepton decays
(\ref{pp_to_ll})-(\ref{nn_to_2nubar}) is
\begin{widetext}
\beq
\Gamma_{NN'} = \Big ( \frac{1}{2m_N} \Big ) \, S \, 
\Big ( \frac{1}{M_{BNV}^{16}} \Big ) \,
\Big (\frac{2}{\pi^3}\Big )^n \, \Big ( \frac{\mu}{M_{BNV}} \Big )^{6n} 
\, 
\Big | \sum_r \bar\kappa^{(NN')}_r e^{-S^{(NN')}_r} 
\langle f.s.| {\cal O}^{(NN')}_r |NN'\rangle \Big |^2 \, R_2 \ , 
\label{gamma_NN_to_ll}
\eeq
\end{widetext}
where $S$ is a symmetry factor, $S=1/2$ for decays with identical leptons in
the final state and $R_2$ is the phase-space factor.

% =======================================================================

\section{ $pp \to \ell^+ \ell'^+$ Decays}
\label{pp_to_ll_section}

In this section we apply our general analysis to the $\Delta L=-2$ dinucleon
decays $pp \to \ell^+ \ell'^+$ of Eq. (\ref{pp_to_ll}), where $\ell$ and
$\ell'$ can be $e$, $\mu$, or $\tau$, as allowed by phase space. Thus, these
are the decays $pp \to (e^+e^+, \mu^+\mu^+, \ e^+\mu^+, \ e^+\tau^+, \ {\rm or}
\ \mu^+\tau^+$).  The $pp \to e^+ e^+$ decay is related by crossing to
hydrogen-anti-hydrogen transitions $(ep) \to (\bar e \bar p)$
\cite{arnellos_marciano}.  These decays are of particular interest because if
an experiment were to observe any of them, this would be not only an
observation of baryon number violation with $\Delta B=-2$, but also an
observation of the violation of total lepton number by $\Delta L=-2$
\cite{majorons}. In contrast, since an experiment does not observe any outgoing
(anti)neutrino(s), the $\Delta L=-2$ decay $np \to \ell^+ \bar\nu$ is
experimentally indistinguishable from the $\Delta L=0$ decay $np \to \ell^+
\nu$.  For the same reason, the $\Delta L=-2$ decay $nn \to \bar\nu \bar \nu'$,
the $\Delta L=0$ decay $nn \to \nu \bar \nu'$, and the $\Delta L=2$ decay $nn
\to \nu \nu'$ are all indistinguishable experimentally.  Furthermore, an
experiment cannot determine whether a final-state neutrino is an EW-doublet
neutrino of some generation ($\nu_e$, $\nu_\mu$, or $\nu_\tau$), or whether it
is an EW-singlet, $\nu_s$.

Because six-quark operators of the form $uud uud$ have nonzero charge
($Q_{em}=2$), they cannot, by themselves, be a singlet under $G_{SM}$.
However, a subset of the six-quark operators is invariant under SU(2)$_L$. The
fact that the six-quark parts of these operators are invariant under SU(2)$_L$
implies that the lepton bilinears must also be invariant under SU(2)$_L$, and
this fixes them to be of the form $[\ell^T_R C \ell'_R]$.  For the set of
we list the following operators, together with the class to which they belong, 
as defined in Table \ref{operator_class_table}:
\begin{widetext}
\beq
{\cal O}^{(pp)}_1 = (T_s)_{\alpha\beta\gamma\delta\rho\sigma}
[u_R^{\alpha \ T} C u_R^\beta]
[u_R^{\gamma \ T} C u_R^\delta]
[d_R^{\rho \ T} C d_R^\sigma][\ell^T_R C \ell'_R]  \quad \in C^{(pp)}_1
\label{op1_pp}
\eeq

\beq
{\cal O}^{(pp)}_2 = (T_s)_{\alpha\beta\gamma\delta\rho\sigma}
[u_R^{\alpha \ T} C d_R^\beta][u_R^{\gamma \ T} C d_R^\delta]
[u_R^{\rho \ T} C u_R^\sigma ][\ell^T_R C \ell'_R] \quad \in C^{(pp)}_1
\label{op2_pp}                  
\eeq
\beq
{\cal O}^{(pp)}_3 = (T_a)_{\alpha\beta\gamma\delta\rho\sigma}
[u_R^{\alpha \ T} C d_R^\beta][u_R^{\gamma \ T} C d_R^\delta]
[u_R^{\rho \ T} C u_R^\sigma ][\ell^T_R C \ell'_R] \quad \in C^{(pp)}_1
\label{op3_pp}                  
\eeq
\beq
{\cal O}^{(pp)}_4 = 
\epsilon_{ij}(T_a)_{\alpha\beta\gamma\delta\rho\sigma}
[Q_L^{i \alpha \ T} C Q_L^{j \beta}][u_R^{\gamma \ T} C d_R^\delta]
[u_R^{\rho \ T} C u_R^\sigma ][\ell^T_R C \ell'_R] \quad \in C^{(pp)}_4
\label{op4_pp}
\eeq
\beq
{\cal O}^{(pp)}_5 = \epsilon_{ij}\epsilon_{km}
(T_a)_{\alpha\beta\gamma\delta\rho\sigma}
[Q_L^{i \alpha \ T} C Q_L^{j \beta}]
[Q_L^{k \gamma \ T} C Q_L^{m \delta}]
[u_R^{\rho \ T} C u_R^\sigma ][\ell^T_R C \ell'_R] \quad \in C^{(pp)}_{10}
\label{op5_pp}
\eeq
and
\beq
{\cal O}^{(pp)}_6 = 
(I_{ss})_{ijkm} (T_s)_{\alpha\beta\gamma\delta\rho\sigma}
[Q_L^{i \alpha \ T} C Q_L^{j \beta}]
[Q_L^{k \gamma \ T} C Q_L^{m \delta}]
[u_R^{\rho \ T} C u_R^\sigma ][\ell^T_R C \ell'_R] \quad \in C^{(pp)}_{10} \ . 
\label{op6_pp}
\eeq
\end{widetext}
The remark concerning linear (in)dependence of operators given above after
Eq. (\ref{op11_nuc3}) also applies here.  There are also operators contributing
to $pp \to \ell^+ \ell'$ in which one or both of the lepton fields is (are)
contained in SU(2)$_L$ doublets rather than being SU(2)$_L$-singlets.  Although
we have carried out an enumeration of these other operations, this enumeration
is actually not necessary for our analysis.  Instead, as before, the key
observation is that the contribution of a given operator ${\cal O}^{(NN')}_r$
to the amplitude for the diproton-to-dilepton decay is determined by the
integrand function (\ref{integral_dinucleon_to_dilepton}), given in general in
Eq. (\ref{integral_dinucleon_to_dilepton}).  Since there are substantially
fewer classes of integrand functions, and hence integrals, than the total
number of operators contributing to $pp \to \ell^+ \ell'^+$, this simplifies
the analysis.  Applying our general formula (\ref{integral_k_gen}), we
calculate the following integrals for the classes of operators contributing to
$pp \to \ell^+ \ell'^+$, as listed in Table \ref{operator_class_table} and Eq.
(\ref{pp_to_ll_classes}).  For the superscript $(NN')$, we list all of the
$\Delta L=-2$ dinucleon decays to which the class contributes. In accord with
our general formula (\ref{integral_k_gen}), we calculate the integrals
\begin{widetext}
\beqs
I^{(pp)}_{C_1} &=& 
b_8 \exp \bigg [ -\frac{1}{8} \Big \{ 
8\| \eta_{u_R} - \eta_{d_R} \|^2 +
4\| \eta_{u_R} - \eta_{\ell_R} \|^2 +
4\| \eta_{u_R} - \eta_{\ell'_R} \|^2 + \cr\cr
&+&
2\| \eta_{d_R} - \eta_{\ell_R} \|^2 +
2\| \eta_{d_R} - \eta_{\ell'_R} \|^2 +
 \| \eta_{\ell_R}-\eta_{\ell'_R} \|^2 \Big \} \bigg ]
\label{integral_class1} 
\eeqs
\beqs
I^{(pp)}_{C_4} &=& b_8 \exp \bigg [ -\frac{1}{8} \Big \{ 
6\| \eta_{Q_L} - \eta_{u_R} \|^2 +
2\| \eta_{Q_L} - \eta_{d_R} \|^2 +
2\| \eta_{Q_L} - \eta_{\ell_R} \|^2 +
2\| \eta_{Q_L} - \eta_{\ell'_R} \|^2 + 
3\| \eta_{u_R} - \eta_{d_R} \|^2 + \cr\cr
&+& 
3\| \eta_{u_R} - \eta_{\ell_R} \|^2 +
3\| \eta_{u_R} - \eta_{\ell'_R} \|^2 + 
 \| \eta_{d_R} - \eta_{\ell_R} \|^2 + 
 \| \eta_{d_R} - \eta_{\ell'_R} \|^2 +
 \| \eta_{\ell_R}-\eta_{\ell'_R} \|^2 \Big \} \bigg ] 
\label{integral_class4} 
\eeqs
\beqs
I^{(pp,np)}_{C_7} &=& b_8 \exp \bigg [ -\frac{1}{8} \Big \{ 
 \| \eta_{Q_L} - \eta_{L_{\ell,L}} \|^2 +
3\| \eta_{Q_L} - \eta_{u_R} \|^2 +
2\| \eta_{Q_L} - \eta_{d_R} \|^2 +
 \| \eta_{Q_L} - \eta_{\ell'_R} \|^2 + 
3\| \eta_{L_{\ell,L}} - \eta_{u_R} \|^2 + \cr\cr
&+& 
2\| \eta_{L_{\ell,L}} - \eta_{d_R} \|^2 +
 \| \eta_{L_{\ell,L}} - \eta_{\ell'_R} \|^2 + 
6\| \eta_{u_R} - \eta_{d_R} \|^2 + 
3\| \eta_{u_R} - \eta_{\ell'_R} \|^2 +
2\| \eta_{d_R}-\eta_{\ell'_R} \|^2 \Big \} \bigg ]
\label{integral_class7} 
\eeqs
\beqs
I^{(pp)}_{C_{10}} &=& 
b_8 \exp \bigg [ -\frac{1}{8} \Big \{ 
8\| \eta_{Q_L} - \eta_{u_R} \|^2 +
4\| \eta_{Q_L} - \eta_{\ell_R} \|^2 +
4\| \eta_{Q_L} - \eta_{\ell'_R} \|^2 + \cr\cr
&+& 
2\| \eta_{u_R} - \eta_{\ell_R} \|^2 + 
2\| \eta_{u_R} - \eta_{\ell'_R} \|^2 +
 \| \eta_{\ell_R} - \eta_{\ell'_R} \|^2  \Big \} \bigg ] 
\label{integral_class10} 
\eeqs
\beqs
I^{(pp,np)}_{C_{13}} &=& b_8 \exp \bigg [ -\frac{1}{8} \Big \{ 
3\| \eta_{Q_L} - \eta_{L_{\ell,L}} \|^2 +
6\| \eta_{Q_L} - \eta_{u_R} \|^2 +
3\| \eta_{Q_L} - \eta_{d_R} \|^2 +
3\| \eta_{Q_L} - \eta_{\ell'_R}\|^2 + 
2\| \eta_{L_{\ell,L}} - \eta_{u_R} \|^2 + \cr\cr
&+&
 \| \eta_{L_{\ell,L}} - \eta_{d_R} \|^2 +
 \| \eta_{L_{\ell,L}} - \eta_{\ell'_R}\|^2 + 
2\| \eta_{u_R} - \eta_{d_R} \|^2 + 
2\| \eta_{u_R} - \eta_{\ell'_R}\|^2 +
 \| \eta_{d_R} - \eta_{\ell'_R} \|^2 \Big \} \bigg ]
\label{integral_class13} 
\eeqs
\beqs
I^{(pp,np,nn)}_{C_{15}} &=& b_8 \exp \bigg [ -\frac{1}{8} \Big \{ 
2\| \eta_{Q_L} - \eta_{L_{\ell,L}} \|^2 +
2\| \eta_{Q_L} - \eta_{L_{\ell',L}}\|^2 +
4\| \eta_{Q_L} - \eta_{u_R} \|^2 +
4\| \eta_{Q_L} - \eta_{d_R} \|^2 +
 \| \eta_{L_{\ell,L}} - \eta_{L_{\ell',L}} \|^2 + \cr\cr
&+&
2\| \eta_{L_{\ell,L}} - \eta_{u_R} \|^2 +
2\| \eta_{L_{\ell,L}} - \eta_{d_R} \|^2 +
2\| \eta_{L_{\ell',L}} -\eta_{u_R} \|^2 +
2\| \eta_{L_{\ell',L}} -\eta_{d_R} \|^2 +
4\| \eta_{u_R} - \eta_{d_R} \|^2 \Big \} \bigg ]
\label{integral_class15}
\eeqs
\beqs
I^{(pp,np,nn)}_{C_{16}} &=& b_8 \exp \bigg [ -\frac{1}{8} \Big \{ 
4\|\eta_{Q_L} - \eta_{L_{\ell,L}} \|^2 +
4\|\eta_{Q_L} - \eta_{L_{\ell',L}} \|^2 +
4\| \eta_{Q_L} - \eta_{u_R} \|^2 + 
4\| \eta_{Q_L} - \eta_{d_R} \|^2 +
 \| \eta_{L_{\ell,L}} - \eta_{L_{\ell',L}} \|^2 + \cr\cr
&+&
 \| \eta_{L_{\ell,L}} - \eta_{u_R} \|^2 +
 \| \eta_{L_{\ell,L}} - \eta_{d_R} \|^2 +
 \| \eta_{L_{\ell',L}} - \eta_{u_R} \|^2 +
 \| \eta_{L_{\ell',L}} - \eta_{d_R} \|^2 +
 \| \eta_{u_R} - \eta_{d_R} \|^2 \Big \} \bigg ] 
\label{integral_class16}
\eeqs
\beqs
I^{(pp,np)}_{C_{17}} &=& b_8 \exp \bigg [ -\frac{1}{8} \Big \{ 
5\| \eta_{Q_L} - \eta_{L_{\ell,L}} \|^2 +
5\| \eta_{Q_L} - \eta_{u_R} \|^2 +
5\| \eta_{Q_L} - \eta_{\ell'_R} \|^2 +
 \| \eta_{L_{\ell,L}} - \eta_{u_R} \|^2 + \cr\cr
&+&
 \| \eta_{L_{\ell,L}} - \eta_{\ell'_R} \|^2 +
 \| \eta_{u_R} - \eta_{\ell'_R} \|^2 \Big \} \bigg ]
\label{integral_class17}
\eeqs
and
\beqs
I^{(pp,np,nn)}_{C_{19}} &=& b_8 \exp \bigg [ -\frac{1}{8} \Big \{ 
6\|\eta_{Q_L} - \eta_{L_{\ell,L}} \|^2 +
6\|\eta_{Q_L} - \eta_{L_{\ell',L}} \|^2 +
 \|\eta_{L_{\ell,L}} - \eta_{L_{\ell',L}} \|^2  \Big \} \bigg ] \ . 
\label{integral_class19}
\eeqs
\end{widetext}

Next, we use the lower bounds on the distances separating the centers of
fermion wavefunctions in the extra dimension that we inferred from lower bounds
on partial lifetimes of proton decay modes.  We substitute these lower bounds
on separation distances into the integrals $I^{(pp)}_n$ and
Eq. (\ref{gamma_NN_to_ll}) to obtain upper bounds on the rates for the $pp \to
\ell^+ \ell'^+$ decays.  Using the lower bounds on the distances separating
centers of fermion wavefunctions that we derived from limits on nucleon decay,
we find that the resultant values of $(\tau/B)_{pp \to \ell^+ \ell'^+} =
(\Gamma_{pp \to \ell^+ \ell'^+})^{-1}$ predicted by the extra-dimensional model
are easily in agreement with current experimental lower bounds on these $\Delta
L=-2$ dinucleon-to-dilepton decays.  As embodied in Eqs. (\ref{ratio_k84}) and
(\ref{log_ratio_k84_example}), this result follows because of the lower bounds
on the exponent sums $S^{(pp)}_r$, together with the fact that the amplitude is
much more highly suppressed, by the prefactor $1/M_{BNV}^8$, as compared with
the prefactor $1/M_{BNV}^2$ that enters in the amplitude for $\Delta L=-1$
nucleon decays such as $p \to \ell^+ \pi^0$, where $M_{BNV}$.  
The lower bounds (from the SK experiment) are \cite{sussman18}:
\beq
(\tau/B)_{pp \to e^+ e^+} > 4.2 \times 10^{33} \ {\rm yr}
\label{tau_pp_to_ee_limit}
\eeq
\beq
(\tau/B)_{pp \to \mu^+ \mu^+} > 4.4 \times 10^{33} \ {\rm yr} 
\label{tau_pp_to_mumu_limit}
\eeq
and
\beq
(\tau/B)_{pp \to e^+ \mu^+} > 4.4 \times 10^{33} \ {\rm yr} 
\label{tau_pp_to_emu_limit}
\eeq
per ${}^{16}O$ nucleus in the water. 

% =======================================================================

\section{$np \to \ell^+ \bar\nu$ Decays}
\label{np_to_lnubar_section}

In this section we proceed to apply the same methods to set upper bounds on
decay rates for the decays $np \to \ell^+ \bar\nu$, where $\ell^+$ can be
$e^+$, $\mu^+$, or $\tau^+$ and $\bar\nu$ can be an electroweak-doublet
antineutrino of any generation or an electroweak-singlet antineutrino.  
Several of the classes of integrals for $np \to \ell^+ \bar\nu$ 
are the same as those for $pp \to \ell^+ \nu$ decays, which we have already
analyzed.  These are the $C^{(NN')}_k$ with $k=7, 13, 15, 16, 17, 19$. 
For the other classes, we calculate the integrals 
\begin{widetext}
\beqs
I^{(np)}_{C_2} &=& b_8 \exp \bigg [ -\frac{1}{8} \Big \{ 
9\| \eta_{u_R} - \eta_{d_R} \|^2 +
3\| \eta_{u_R} - \eta_{\ell_R} \|^2 +
3\| \eta_{u_R} - \eta_{\nu_{s,R}} \|^2 +
3\| \eta_{d_R} - \eta_{\ell_R} \|^2 + \cr\cr
&+&
3 \| \eta_{d_R} - \eta_{\nu_{s,R}} \|^2 +
 \| \eta_{\ell_R} - \eta_{\nu_{s,R}} \|^2 \Big \} \bigg ]
\label{integral_class2}
\eeqs
\beqs
I^{(np)}_{C_5} &=& b_8 \exp \bigg [ -\frac{1}{8} \Big \{ 
4\| \eta_{Q_L} - \eta_{u_R} \|^2 +
4\| \eta_{Q_L} - \eta_{d_R} \|^2 +
2\| \eta_{Q_L} - \eta_{\ell_R} \|^2 + 
2\| \eta_{Q_L} - \eta_{\nu_{s,R}} \|^2 + 
4\| \eta_{u_R} - \eta_{d_R} \|^2 + \cr\cr
&+&
2\| \eta_{u_R} - \eta_{\ell_R} \|^2 +
2\| \eta_{u_R} - \eta_{\nu_{s,R}} \|^2 +
2\| \eta_{d_R} - \eta_{\ell_R} \|^2 +
2\| \eta_{d_R} - \eta_{\nu_{s,R}} \|^2 +
 \| \eta_{\ell_R} - \eta_{\nu_{s,R}} \|^2 \Big \} \bigg ]
\label{integral_class5}
\eeqs
\beqs
I^{(np,nn)}_{C_8} &=& b_8 \exp \bigg [ -\frac{1}{8} \Big \{ 
 \| \eta_{Q_L} - \eta_{L_{\ell,L}} \|^2 +
2\| \eta_{Q_L} - \eta_{u_R} \|^2 +
3\| \eta_{Q_L} - \eta_{d_R} \|^2 + 
 \| \eta_{Q_L} - \eta_{\nu_{s,R}} \|^2 + 
2\| \eta_{L_{\ell,L}} - \eta_{u_R} \|^2 + \cr\cr
&+&
3\| \eta_{L_{\ell,L}} - \eta_{d_R} \|^2 +
 \| \eta_{L_{\ell,L}} - \eta_{\nu_{s,R}} \|^2 +
6\| \eta_{u_R} - \eta_{d_R} \|^2 +
2\| \eta_{u_R} - \eta_{\nu_{s,R}} \|^2 + 
3\| \eta_{d_R} - \eta_{\nu_{s,R}} \|^2 \Big \} \bigg ]
\label{integral_class8}
\eeqs
\beqs
I^{(np)}_{C_9} &=& b_8 \exp \bigg [ -\frac{1}{8} \Big \{ 
 \| \eta_{L_{\ell,L}} - \eta_{L_{\ell',L}} \|^2 +
3\| \eta_{L_{\ell,L}} - \eta_{u_R} \|^2 +
3\| \eta_{L_{\ell,L}} - \eta_{d_R} \|^2 +
3\| \eta_{L_{\ell',L}} - \eta_{u_R} \|^2 + \cr\cr
&+&
3\| \eta_{L_{\ell',L}} - \eta_{d_R} \|^2 +
9\| \eta_{u_R} - \eta_{d_R} \|^2 \Big \} \bigg ]
\label{integral_class9}
\eeqs
\beqs
I^{(np)}_{C_{11}} &=& b_8 \exp \bigg [ -\frac{1}{8} \Big \{ 
4\|\eta_{Q_L} - \eta_{u_R} \|^2 +
4\|\eta_{Q_L} - \eta_{d_R} \|^2 +
4\| \eta_{Q_L} - \eta_{\ell_R} \|^2 + 
4\| \eta_{Q_L} - \eta_{\nu_{s,R}} \|^2 + 
 \| \eta_{u_R} - \eta_{d_R} \|^2 + \cr\cr
&+&
 \| \eta_{u_R} - \eta_{\ell_R} \|^2 +
 \| \eta_{u_R} - \eta_{\nu_{s,R}} \|^2 +
 \| \eta_{d_R} - \eta_{\ell_R} \|^2 +
 \| \eta_{d_R} - \eta_{\nu_{s,R}} \|^2 + 
 \| \eta_{\ell_R} - \eta_{\nu_{s,R}} \|^2 \Big \} \bigg ]
\label{integral_class11}
\eeqs
\beqs
I^{(np,nn)}_{C_{14}} &=& b_8 \exp \bigg [ -\frac{1}{8} \Big \{ 
3\| \eta_{Q_L} - \eta_{L_{\ell,L}} \|^2 +
3\| \eta_{Q_L} - \eta_{u_R} \|^2 +
6\| \eta_{Q_L} - \eta_{d_R} \|^2 + 
3\| \eta_{Q_L} - \eta_{\nu_{s,R}} \|^2 + 
 \| \eta_{L_{\ell,L}} - \eta_{u_R} \|^2 + \cr\cr
&+&
2\| \eta_{L_{\ell,L}} - \eta_{d_R} \|^2 +
 \| \eta_{L_{\ell,L}} - \eta_{\nu_{s,R}} \|^2 +
2\| \eta_{u_R} - \eta_{d_R} \|^2 +
 \| \eta_{u_R} - \eta_{\nu_{s,R}} \|^2 + 
2\| \eta_{d_R} - \eta_{\nu_{s,R}} \|^2 \Big \} \bigg ]
\label{integral_class14}
\eeqs
and
\beqs
I^{(np,nn)}_{C_{18}} &=& b_8 \exp \bigg [ -\frac{1}{8} \Big \{ 
5\| \eta_{Q_L} - \eta_{L_{\ell,L}} \|^2 +
5\| \eta_{Q_L} - \eta_{d_R} \|^2 +
5\| \eta_{Q_L} - \eta_{\nu_{s,R}} \|^2 +
 \| \eta_{L_{\ell,L}} - \eta_{d_R} \|^2 + \cr\cr
&+&
 \| \eta_{L_{\ell,L}} - \eta_{\nu_{s,R}} \|^2 +
 \| \eta_{d_R} -        \eta_{\nu_{s_R}} \|^2 \Big \} \bigg ] \ . 
\label{integral_class18}
\eeqs
\end{widetext} 
Although it is not necessary for our analysis, one can construct explicit
operators of each class, as we have done for the operators contributing to $pp
\to \ell^+ \ell'^+$.  Some of these contribute to decays with EW-singlet
antineutrinos, while others contribute to decays with EW-doublet antineutrinos,
but since these decays are indistinguishable experimentally, we include all of
these operators together. For example, there are several operators in which all
fermions are SU(2)$_L$ singlets: 
\begin{widetext}
\beq
{\cal O}^{(np)}_1 = (T_s)_{\alpha\beta\gamma\delta\rho\sigma}
[u_R^{\alpha \ T} C u_R^\beta]
[u_R^{\gamma \ T} C d_R^\delta]
[d_R^{\rho \ T} C d_R^\sigma][\ell^T_R C \nu_{s,R}] \quad \in C^{(np)}_2
\label{op1_np}
\eeq
\beq
{\cal O}^{(np)}_2 = (T_s)_{\alpha\beta\gamma\delta\rho\sigma}
[u_R^{\alpha \ T} C d_R^\beta][u_R^{\gamma \ T} C d_R^\delta]
[u_R^{\rho \ T} C d_R^\sigma ][\ell^T_R C \nu_{s,R}] \quad \in C^{(np)}_2 
\label{op2_np}
\eeq
\beq
{\cal O}^{(np)}_3 = (T_a)_{\alpha\beta\gamma\delta\rho\sigma}
[u_R^{\alpha \ T} C d_R^\beta][u_R^{\gamma \ T} C d_R^\delta]
[u_R^{\rho \ T} C d_R^\sigma ][\ell^T_R C \nu_{s,R}] \quad \in C^{(np)}_2 
\label{op3_np}
\eeq
\beq
{\cal O}^{(np)}_4 =
\epsilon_{ij}(T_a)_{\alpha\beta\gamma\delta\rho\sigma}
[Q_L^{i \alpha \ T} C Q_L^{j \beta}][u_R^{\gamma \ T} C d_R^\delta]
[u_R^{\rho \ T} C d_R^\sigma ][\ell^T_R C \nu_{s,R}] \quad \in C^{(np)}_5 
\label{op4_np}
\eeq
\beq
{\cal O}^{(np)}_5 = \epsilon_{ij}\epsilon_{km}
(T_a)_{\alpha\beta\gamma\delta\rho\sigma}
[Q_L^{i \alpha \ T} C Q_L^{j \beta}]
[Q_L^{k \gamma \ T} C Q_L^{m \delta}]
[u_R^{\rho \ T} C d_R^\sigma ][\ell^T_R C \nu_{s,R}] \quad \in C^{(np)}_{11} 
\label{op5_np}
\eeq
and
\beq
{\cal O}^{(np)}_6 =
(I_{ss})_{ijkm} (T_s)_{\alpha\beta\gamma\delta\rho\sigma}
[Q_L^{i \alpha \ T} C Q_L^{j \beta}]
[Q_L^{k \gamma \ T} C Q_L^{m \delta}]
[u_R^{\rho \ T} C d_R^\sigma ][\ell^T_R C \nu_{s,R}] \quad \in C^{(np)}_{11} 
\ . 
\label{op6_np}
\eeq
\end{widetext}
There are also operators contributing to $np \to \ell^+ \bar\nu$ in
which one or both of the lepton fields is (are) contained in SU(2)$_L$ doublets
rather than being SU(2)$_L$-singlets.  We have constructed these explicitly,
using the same methods that we used for the corresponding operators
contributing to $pp \to \ell^+ \ell'^+$. 

Proceeding as in Sec. \ref{pp_to_ll_section}, we have calculated the resultant
rates for the $\Delta L =-2$ decays $np \to \ell^+ \bar\nu$.  Using the lower
bounds on distances between fermion wavefunction centers in the extra
dimensions that we have derived in Section \ref{pdecay_section}, we find that
the resultant lower bounds on the partial lifetimes are in agreement with the
current experimental lower bounds on these decays.  Furthermore,
as noted earlier, since an experiment would not observe the outgoing
antineutrino, it would not be able to distinguish the $\Delta L=-2$ decay $np
\to \ell^+ \bar\nu$ from the $\Delta L=0$ decay $np \to \ell^+ \nu$.  As
discussed in \cite{dnd}, the latter decay can occur via the combination of a
six-quark BNV vertex with SM fermion processes and hence is generically much
less suppressed than the $\Delta L=-2$ dinucleon-to-dilepton decays.

% ======================================================================

\section{$nn \to \bar\nu \bar\nu'$ and $nn \to \nu \nu'$ Decays}
\label{nn_to_2nubar_section}

In this section we consider the $\Delta L=-2$ dineutron decay 
$nn \to \bar\nu \bar\nu'$ and the corresponding $\Delta L=2$ decay
$nn \to \nu \nu'$. Of the classes of eight-fermion operators contributing to the
$\Delta L=-2$ dineutron decay $nn \to \bar\nu \bar\nu'$, the six resultant
$I^{(NN')}_k$ integrals have already been given above, namely those for
$k=8, \ 14, \  15, \ 16, \ 18$, and 19. The remaining three integrals are for
$k=3, \ 6, \ 12$. We calculate the integrals 
\begin{widetext}
\beqs
I^{(nn)}_{C_3} &=& b_8 \exp \bigg [ -\frac{1}{8} \Big \{
8\| \eta_{u_R} - \eta_{d_R} \|^2 +
2\| \eta_{u_R} - \eta_{\nu_{s,R}} \|^2 +
2\| \eta_{u_R} - \eta_{\nu_{s',R}} \|^2 +
4\| \eta_{d_R} - \eta_{\nu_{s,R}} \|^2 + \cr\cr
&+&
4\| \eta_{d_R} - \eta_{\nu_{s',R}} \|^2 +
 \| \eta_{\nu_{s,R}} - \eta_{\nu_{s',R}} \|^2 \Big \} \bigg ]
 \label{integral_class3}
\eeqs
\beqs
I^{(nn)}_{C_6} &=& b_8 \exp \bigg [ -\frac{1}{8} \Big \{
2\| \eta_{Q_L} - \eta_{u_R} \|^2 +
6\| \eta_{Q_L} - \eta_{d_R} \|^2 +
2\| \eta_{Q_L} - \eta_{\nu_{s,R}} \|^2 +
2\| \eta_{Q_L} - \eta_{\nu_{s',R}} \|^2 + 
3\| \eta_{u_R} - \eta_{d_R} \|^2 + \cr\cr
&+&
 \| \eta_{u_R} - \eta_{\nu_{s,R}} \|^2 +
 \| \eta_{u_R} - \eta_{\nu_{s',R}} \|^2 +
3\| \eta_{d_R} - \eta_{\nu_{s,R}} \|^2 +
3\| \eta_{d_R} - \eta_{\nu_{s',R}} \|^2 + 
 \| \eta_{\nu_{s,R}} - \eta_{\nu_{s',R}} \|^2 \Big \} \bigg ]
\label{integral_class6}
\eeqs
and
\beqs
I^{(nn)}_{C_{12}} &=& b_8 \exp \bigg [ -\frac{1}{8} \Big \{
8\| \eta_{Q_L} - \eta_{d_R} \|^2 +
4\| \eta_{Q_L} - \eta_{\nu_{s,R}} \|^2 +
4\| \eta_{Q_L} - \eta_{\nu_{s',R}} \|^2 +
2\| \eta_{d_R} - \eta_{\nu_{s,R}} \|^2 + \cr\cr
&+&
2\| \eta_{d_R} - \eta_{\nu_{s',R}} \|^2 +
 \| \eta_{\nu_{s,R}} - \eta_{\nu_{s',R}} \|^2 \Big \} \bigg ] \ .
\label{integral_class12}
\eeqs
\end{widetext}
Applying our lower bounds on the distances between centers of fermion
wavefunctions in the extra dimension from Section \ref{pdecay_section}, we find
that these $\Delta L=-2$ dinucleon decays are highly suppressed, similar to
what we showed for the $pp \to \ell^+ \ell'^+$ and $np \to \ell^+ \bar\nu$
decays. 

One can also consider the $\Delta L=2$ dineutron-to-dilepton decays $nn \to \nu
\nu'$ in Eq. (\ref{nn_to_2nu}).  Given that $\nu_{s,R}$ is assigned lepton
number $L=1$, there is a corresponding charge-conjugate field, $(\nu_{s,R})^c =
(\nu^c_s)_L$ with lepton number $L=-1$.  The eight-fermion operators that
contribute to the decays (\ref{nn_to_2nu}) are obtained from those for the
decay $nn \to \bar\nu \bar\nu'$ by replacing the $[\nu_{s,R}^T C \nu_{s',R}]$
neutrino bilinear by $[(\nu^c_s)^T_L C (\nu^c_{s'})_L]$. There are thus three
classes of operators, which are the results of this change applied to the
classes $C^{(nn)}_k$ with $k=3, \ 6, \ 12$ for $nn \to \bar\nu \bar\nu'$
decays.  Carrying out the resultant analysis, we reach the same conclusions as
we did for the $\Delta L=-2$ dinucleon-to-dilepton decays concerning the highly
suppressed rates.

A general comment concerning both of these $\Delta L = \pm 2$ 
dineutron decays is that since an experiment would not observe the outgoing
(anti)neutrinos, it could not distinguish these decays from the $\Delta L=0$
dineutron decays $nn \to \nu \bar\nu$ decays, which 
can occur via a six-quark BNV operator combined
with SM processes and hence are generically much less suppressed than the 
$\Delta L=-2$ decays $nn \to \bar\nu \bar\nu'$ \cite{dnd}.  

One also expects similar suppression in this extra-dimensional model for $B$-
and $L$-violating decays involving trinucleon initial states, such as $ppp \to
\ell^+ \pi^+ \pi^+$ and $ppn \to \ell^+ \pi^+$, mediated by 10-fermion
operators, or $ppp \to \ell^+ \ell'^+ \ell''^+$ and $ppn \to \ell^+ \ell'^+
\bar\nu$, mediated by 12-fermion operators. Recent experimental bounds on
trinucleon decays include \cite{exo200,maj}.

% =====================================================================

\section{Conclusions}
\label{conclusion_section} 

In this paper we have studied several baryon-number-violating nucleon and
dinucleon decays in a model with large extra dimensions, including (i) the
$\Delta L=-3$ nucleon decays $p \to \ell^+ \bar\nu \bar\nu'$ and $n \to \bar\nu
\bar\nu' \bar\nu''$; (ii) the $\Delta L = 1$ nucleon decays $p \to \ell^+ \nu
\nu'$ and $n \to \bar\nu \nu' \nu''$; (iii) the $\Delta L=-2$ dinucleon decays
$pp \to (e^+e^+, \mu^+\mu^+, \ e^+\mu^+, \ e^+\tau^+, \ {\rm or} \
\mu^+\tau^+$), $np \to \ell^+ \bar\nu$, and $nn \to \bar\nu \bar\nu'$, where
$\ell^+=e^+, \ \mu^+$, or $\tau^+$; and (iv) the $\Delta L=2$ dineutron decays
$nn \to \nu \nu'$.  The decays of type (i) and (ii) are mediated by six-fermion
operators, while the decays of type (iii) and (iv) are mediated by
eight-fermion operators. Motivated by the earlier finding in Ref. \cite{nnb02}
that, even with fermion wavefunction positions chosen so as to render the rates
for baryon-violating nucleon decays much smaller than experimental limits,
$n-\bar n$ oscillations could occur at rates comparable to experimental bounds,
we have addressed the generalized question of whether nucleon and dinucleon
decays to leptonic final states mediated by six-fermion and eight-fermion
operators are sufficiently suppressed to agree with experimental bounds.  To
investigate this question, we have determined constraints on separations
between wavefunctions in the extra dimensions from limits on the best
constrainted proton and bound neutron decay modes, and then have applied these
in analyses of relevant six-fermion and eight-fermion operators contributing to
the decays (i)-(iv).  From these analyses, we find that in this
extra-dimensional model, these decays are strongly suppressed, in accord with
experimental limits. The reason that $n-\bar n$ oscillations can occur at a
level comparable with current limits, while the decays (i)-(iv) are suppressed
well below experimental limits on the respective modes can be traced to the
fact that nucleon decays can be suppressed by making the separations between
quark and lepton wavefunction centers sufficiently large. This procedure does
not suppress $n-\bar n$ oscillations, but considerably suppresses the
baryon-violating decays of nucleons and dinucleons considered here. In addition
to its phenomenological value, our analysis provides an interesting example of
the application of low-energy effective field theory techniques to a problem
involving several relevant mass scales. Here, these mass scales include the
fermion wavefunction localization parameter $\mu$, the overall mass scale of
baryon number violation, $M_{BNV}$, and the multiple inverse separation
distances $\|y_{f_i} - y_{f_j} \|^{-1}$ between various fermion wavefunction
centers in the extra dimensions.

% =======================================================================

\begin{acknowledgments}

This research was supported in part by the NSF Grants NSF-PHY-1620628 and 
NSF-PHY-1915093 (R.S.). 

\end{acknowledgments}

% =======================================================================

\begin{appendix}

% =======================================================================

\section{Some Integrals}
\label{integral_appendix}

We record here some relevant formulas that are used for our
calculations. First, with $\eta$ a real variable and the (real) constants 
$a_i > 0$, $i=1,...,n$, we have 
\begin{widetext}
\beq
\int_{-\infty}^\infty d \eta \, \exp \Big [
-\sum_{i=1}^n a_i \, (\eta-\eta_{f_i})^2 \Big ]  
= \bigg [ \frac{\pi}{\sum_{i=1}^n a_i} \bigg ]^{1/2} \, 
\exp\Bigg [ \frac{-\sum_{j,k=1; \ j < k}^n  \, a_j a_k \, 
(\eta_{f_j}-\eta_{f_k})^2}{\sum_{s=1}^n a_s} \Bigg ] \ . 
\label{intform1}
\eeq
The sum $\sum_{j,k=1; \ j < k}^n  \, a_j a_k \ (\eta_{f_j}-\eta_{f_k})^2$
contains ${n \choose 2}$ terms, where ${n \choose m} \equiv n!/[m! (n-m)!]$
is the binomial coefficient. 

Second, now generalizing $\eta$ to an $n$-dimensional vector $\eta \in
{\mathbb R}^n$ with components $\eta_j$, $j=1,...,n$ and 
norm $\| \eta \| = [\sum_{j=1}^n \eta_j^2]^{n/2}$, and denoting 
$[\prod_{j=1}^n \int_{-\infty}^\infty d\eta_j]\, F(\eta) 
\equiv \int d^n \eta \, F(\eta)$, we have 
\beq
\int d^n \eta \, \exp
\Big [-\sum_{i=1}^n a_i\|\eta-\eta_{f_i}\|^2 \Big ]
= \bigg [ \frac{\pi}{\sum_{i=1}^n a_i} \bigg ]^{n/2} \,
\exp\Bigg [ \frac{-\sum_{j,k=1; \ j < k}^n \, a_j a_k
\|\eta_{f_j}-\eta_{f_k}\|^2}{\sum_{s=1}^n a_s} \Bigg ] \ .
\label{intform}
\eeq
Thus, for example, for $n=3$, 
\beqs
&& \int d^n \eta \, \exp \Big [- \Big (
a_1\|\eta-\eta_{f_1}\|^2 +
a_2\|\eta-\eta_{f_2}\|^2 +
a_3\|\eta-\eta_{f_3}\|^2 \Big ) \Big ] = \cr\cr
&=& \bigg [ \frac{\pi}{a_1+a_2+a_3} \bigg ]^{n/2} \, \exp \Bigg [ \frac{-
\Big (
a_1a_2\|\eta_{f_1}-\eta_{f_2}\|^2+
a_2a_3\|\eta_{f_2}-\eta_{f_3}\|^2+
a_3a_1\|\eta_{f_3}-\eta_{f_1}\|^2 \Big )}{a_1+a_2+a_3} \Bigg ] \ . 
\label{intform_ex}
\eeqs
\end{widetext}
%

% =======================================================================

\section{Properties of Color Tensors}
\label{colortensor_appendix}

The tensors $T_s$ and $T_a$ in Eqs. (\ref{ts}) and (\ref{ta}) were defined and
used in \cite{nnb82}; in \cite{nnb84} their properties were discussed further
and a third type of color tensor, denoted $T_{a3}$, was defined and applied. 
In this appendix we review the properties of these tensors. We use the 
notation $(a,b)$ and $[a,b]$to mean, respectively, symmetry and antisymmetry
under the interchange $a \leftrightarrow b$, where $a$ and $b$ can be single
SU(3)$_c$ indices or sets of indices.  The tensor $T_s$ has the properties 
\beqs
(T_s)_{\alpha \beta \gamma \delta \rho \sigma}: \ && (\alpha,\beta), \quad 
(\gamma,\delta), \quad (\rho,\sigma), \cr\cr
&& (\alpha\beta,\gamma\delta), \quad (\gamma\delta,\rho\sigma), \quad
 (\alpha\beta,\rho\sigma) \ .  \cr\cr
&&  
\label{ts_properties}
\eeqs
Thus, in a contraction of $T_s$ with a product of six fundamental
(\underline{3}) representations of SU(3)$_c$, the first two pairs are each
combined as $(3 \times 3)_s = 6$, i.e., in terms of Young tableaux, 
$(\fund \times \fund)_s = \sym$; then the resultant two \underline{6}
representations are combined symmetrically as $(6 \times 6)_s = \bar 6$, 
i.e., $(\sym \times \sym)_s = \overline{\sym}$, 
and finally this $\bar 6$ is combined with the 6 resulting from the third 
pair $(3 \times 3)_s = 6$ to make an SU(3)$_c$ singlet.

The tensor $T_a$ has the properties
\beq
(T_a)_{\alpha \beta \gamma \delta \rho \sigma}: \ [\alpha,\beta], \quad 
[\gamma,\delta], \quad (\rho,\sigma), \quad (\alpha\beta,\gamma\delta) \ . 
\label{ta_properties}
\eeq
Hence, in a contraction of $T_a$ with a product of six fundamental
representations of SU(3)$_c$, the first two pairs are each combined as $(3
\times 3)_a = \bar 3$, then the resultant two $\bar 3$ representations are
combined as $(\bar 3 \times \bar 3)_s = \bar 6$, and finally, this is combined
with the 6 from the $(\rho\sigma)$ combination to make an SU(3)$_c$ singlet.
To indicate more explicitly these (anti)symmetry properties, Ref. \cite{nnb84}
introduced the notation
\beq
(T_a)_{\alpha\beta\gamma\delta\rho\sigma} \equiv 
(T_{aas})_{\alpha\beta\gamma\delta\rho\sigma} \ , 
\label{taas}
\eeq
where the subscript $(aas)$ refers to the antisymmetry on the first two pairs
of color indices and symmetry on the last pair.  In an obvious notation, there 
are two other related color tensors, $T_{asa}$ and $T_{saa}$. 

As noted in \cite{nnb84}, there is a third way to couple six fundamental
representations of SU(3)$_c$ together to make a singlet, namely to couple each
pair antisymmetrically, via the tensor $T_{a3}$ was given in Eq. (\ref{ta3}).
This tensor was not needed in the analysis of $n-\bar n$ oscillations in
\cite{nnb82} but did enter in the analysis of six-quark operators involving
higher generations in \cite{nnb84}.  It has the properties
\beqs
(T_{a3})_{\alpha\beta\gamma\delta\rho\sigma}: \ && [\alpha,\beta], \quad 
[\gamma,\delta], \quad [\rho,\sigma], \cr\cr
&& [\alpha\beta,\gamma\delta], \quad [\gamma\delta,\rho\sigma], \quad
   [\alpha\beta,\rho\sigma] \ . \cr\cr
&& 
\label{ta3_properties}
\eeqs
%

% =========================================================================

\section{Phase Space Factors}
\label{phase_space_appendix}

For an initial state with invariant mass $\sqrt{s}$ decaying to an $n$-body 
final state $f.s.$, the phase space factor is 
\beq
\int dR_n = \frac{1}{(2\pi)^{3n-4}} \,
\int \Big [ \prod_{i=1}^n
\frac{d^3 p_i}{2E_i} \Big ] \, \delta^4 \Big ( p-(\sum_{i=1}^n p_i) \Big ) \ .
\label{phase_space_integral}
\eeq
where $p$ is the four-momentum of the initial state, and $E_i$ and $p_i$ 
denote the energies and four-momenta of the final-state particles. 
We define the Lorenz-invariant phase space factor as 
\beq
R_n = \int dR_n \ . 
\label{nbody_phasespace}
\eeq
We will only need $R_2$, which is 
\beq
R_2 = \frac{1}{8\pi} \, [\lambda(1,\delta_1,\delta_2)]^{1/2} \ , 
\label{r2}
\eeq
where $\lambda(x,y,z) = x^2+y^2+z^2-2(xy+yz+zx)$ and $\delta_i = m_i^2/s$.  If
$m_i^2/s$ is zero or negligibly small for all particles $i$ in the final state,
then $R_2=1/(8\pi)$. If $\delta_1=\delta_2 \equiv \delta$, then
$R_2=(8\pi)^{-1}\sqrt{1-4\delta}$. 

% ======================================================================

\section{Operators Contributing to $pp \to \ell^+ \ell'^+$}
\label{pp_operator_appendix}

Although our results in this paper depend only on the classes of operators
$C^{(NN')}_k$ and the resultant integrals of fermion fields over the extra
dimensions, $I^{(NN')}_{C_k} \equiv I_{C^{(NN')}_k}$, it is worthwhile, for
illustrative purposes, to display various explicit operators that contribute to
the $\Delta L=-2$ diproton decays $pp \to \ell^+ \ell'^+$.  We have listed
operators of this type in which all fermions are SU(2)$_L$ singlets in the
text.  Here we give operators contributing to $pp \to \ell^+ \ell'^+$ in which
one or both of the lepton fields is (are) in SU(2)$_L$ doublets.  As remarked
after Eq. (\ref{op11_nuc3}) in the text, since our analysis only depends on the
classes of operators (defined by the integrals), which are manifestly
independent, since they are comprised of different fermion fields, it is not
necessary to work out all linear independence properties among these explicit
operators.

Operators with one lepton field arising from an SU(2)$_L$ doublet and the other
an SU(2)$_L$ singlet include the following. The first of these is 
\begin{widetext}
\beq
{\cal O}^{(pp,np)}_7 =
\epsilon_{ij}(T_s)_{\alpha\beta\gamma\delta\rho\sigma}
[Q_L^{i \alpha \ T} C L^j_{\ell,L}][u^{\beta \ T}_R C d^\gamma_R]
[u_R^{\delta \ T} C u_R^\rho ][d^{\sigma \ T}_R C \ell'_R] \ 
\in C^{(pp,np)}_7 \ .
\label{op7_pp}
\eeq
Carrying out the SU(2)$_L$ contractions in ${\cal
 O}^{(pp,np)}_7$ explicitly, one has
\beq
{\cal O}^{(pp,np)}_7 = (T_s)_{\alpha\beta\gamma\delta\rho\sigma}
\Big ( [u^{\alpha \ T}_L C \ell_L] - [d^{\alpha \ T}_L C \nu_{\ell,L}] \Big )
[u^{\beta \ T}_R C d^\gamma_R]
[u_R^{\delta \ T} C u_R^\rho ][d^{\sigma \ T}_R C \ell'_R] \ .
\label{op7_pp_explicit}
\eeq
Of the two terms in Eq. (\ref{op7_pp_explicit}), the one containing the
$[u^{\alpha \ T}_L C \ell_L]$ fermion bilinear contributes to $pp \to \ell^+
\ell'^+$, while the other term contributes to $np \to \ell'^+ \bar \nu_\ell$. 
Since it is straightforward to determine which dinucleon-to-dilepton
decays each operator contributes to, we do not indicate this explicitly. 
Other operators include
\beq
{\cal O}^{(pp,np)}_8 = 
\epsilon_{ij}(T_s)_{\alpha\beta\gamma\delta\rho\sigma}
[Q_L^{i \alpha \ T} C L^j_{\ell,L}][u^{\beta \ T}_R C d^\gamma_R]
[u_R^{\delta \ T} C d_R^\rho ][u^{\sigma \ T}_R C \ell'_R] \ 
\in C^{(pp,np)}_7 
\label{op8_pp}
\eeq
\beq
{\cal O}^{(pp,np)}_9 = 
\epsilon_{ij}(T_s)_{\alpha\beta\gamma\delta\rho\sigma}
[Q_L^{i \alpha \ T} C L^j_{\ell,L}][u^{\beta \ T}_R C u^\gamma_R]
[d_R^{\delta \ T} C d_R^\rho ][u^{\sigma \ T}_R C \ell'_R] \ \in C^{(pp,np)}_7 
\label{op9_pp}
\eeq
\beq
{\cal O}^{(pp,np)}_{10} = 
\epsilon_{ij}(T_a)_{\alpha\beta\gamma\delta\rho\sigma}
[Q_L^{i \alpha \ T} C L^j_{\ell,L}][u^{\beta \ T}_R C d^\gamma_R]
[u_R^{\delta \ T} C u_R^\rho ][d^{\sigma \ T}_R C \ell'_R] \ \in C^{(pp,np)}_7 
\label{op10_pp}
\eeq
\beq
{\cal O}^{(pp,np)}_{11} = 
\epsilon_{ij}(T_a)_{\alpha\beta\gamma\delta\rho\sigma}
[Q_L^{i \alpha \ T} C L^j_{\ell,L}][u^{\beta \ T}_R C d^\gamma_R]
[u_R^{\delta \ T} C d_R^\rho ][u^{\sigma \ T}_R C \ell'_R] \ \in C^{(pp,np)}_7 
\label{op11_pp}
\eeq
\beq
{\cal O}^{(pp,np)}_{12} = 
\epsilon_{ij}\epsilon_{km}(T_a)_{\alpha\beta\gamma\delta\rho\sigma}
[Q_L^{i \alpha \ T} C Q^{j \beta}_L][Q^{k \gamma \ T}_L C L^m_{\ell,L}]
[u_R^{\rho \ T} C u_R^\sigma ][d^{\delta \ T}_R C \ell'_R] \ \in 
C^{(pp,np)}_{13} 
\label{op12_pp}
\eeq
\beq
{\cal O}^{(pp,np)}_{13} = 
\epsilon_{ij}\epsilon_{km}(T_a)_{\alpha\beta\gamma\delta\rho\sigma}
[Q_L^{i \alpha \ T} C Q^{j \beta}_L][Q^{k \gamma \ T}_L C L^m_{\ell,L}]
[u_R^{\rho \ T} C d_R^\sigma ][u^{\delta \ T}_R C \ell'_R] \ \in 
C^{(pp,np)}_{13} 
\label{op13_pp}
\eeq
\beq
{\cal O}^{(pp,np)}_{14} = 
\epsilon_{ij}\epsilon_{km}(T_{a3})_{\alpha\beta\gamma\delta\rho\sigma}
[Q_L^{i \alpha \ T} C Q^{j \beta}_L][Q^{k \gamma \ T}_L C L^m_{\ell,L}]
[u_R^{\rho \ T} C d_R^\sigma ][u^{\delta \ T}_R C \ell'_R] \ \in 
C^{(pp,np)}_{13} 
\label{op14_pp}
\eeq
\beq
{\cal O}^{(pp,np)}_{15} = 
\epsilon_{ij}\epsilon_{km}\epsilon_{np}
(T_a)_{\alpha\beta\gamma\delta\rho\sigma}
[Q_L^{i \alpha \ T} C Q^{j \beta}_L][Q_L^{k \gamma \ T} C Q^{m \delta}_L]
[Q^{n \rho \ T}_L C L^p_{\ell,L}][u^{\sigma \ T}_R C \ell'_R] \ \in 
C^{(pp,np)}_{17} 
\label{op15_pp}
\eeq
\beq
{\cal O}^{(pp,np)}_{16} = 
(I_{ss})_{ijkm}(T_s)_{\alpha\beta\gamma\delta\rho\sigma}
[Q_L^{i \alpha \ T} C Q^{j \beta}_L][Q^{k \gamma \ T}_L C L^m_{\ell,L}]
[u_R^{\rho \ T} C u_R^\sigma ][d^{\delta \ T}_R C \ell'_R] \ \in 
C^{(pp,np)}_{13}
\label{op16_pp}
\eeq
\beq
{\cal O}^{(pp,np)}_{17} = 
(I_{ss})_{ijkm}(T_s)_{\alpha\beta\gamma\delta\rho\sigma}
[Q_L^{i \alpha \ T} C Q^{j \beta}_L][Q^{k \gamma \ T}_L C L^m_{\ell,L}]
[u_R^{\rho \ T} C d_R^\sigma ][u^{\delta \ T}_R C \ell'_R] \ \in 
C^{(pp,np)}_{13} 
\label{op17_pp}
\eeq
\beq
{\cal O}^{(pp,np,nn)}_{18} = 
\epsilon_{ij}\epsilon_{km}(T_s)_{\alpha\beta\gamma\delta\rho\sigma}
[Q_L^{i \alpha \ T} C L^j_{\ell,L}][Q_L^{k \beta \ T} C L^m_{\ell',L}]
[u_R^{\gamma \ T} C u_R^\delta ][d_R^{\rho \ T} C d_R^\sigma ] \ \in 
C^{(pp,np,nn)}_{15} 
\label{op18_pp}
\eeq
\beq
{\cal O}^{(pp,np,nn)}_{19} = 
\epsilon_{ij}\epsilon_{km}(T_s)_{\alpha\beta\gamma\delta\rho\sigma}
[Q_L^{i \alpha \ T} C L^j_{\ell,L}][Q_L^{k \beta \ T} C L^m_{\ell',L}]
[u_R^{\gamma \ T} C d_R^\delta ][u_R^{\rho \ T} C d_R^\sigma ] \ \in 
C^{(pp,np,nn)}_{15} 
\label{op19_pp}
\eeq
\beq
{\cal O}^{(pp,np,nn)}_{20} = 
\epsilon_{ij}\epsilon_{km}(T_a)_{\alpha\beta\gamma\delta\rho\sigma}
[Q_L^{i \rho \ T} C L^j_{\ell,L}][Q_L^{k \sigma \ T} C L^m_{\ell',L}]
[u_R^{\alpha \ T} C d_R^\beta ][u_R^{\ \gamma T} C d_R^\delta ] \ \in 
C^{(pp,np,nn)}_{15} 
\label{op20_pp}
\eeq 
\beq
{\cal O}^{(pp,np,nn)}_{21} = 
\epsilon_{ij}\epsilon_{km}\epsilon_{np}
(T_a)_{\alpha\beta\gamma\delta\rho\sigma}
[Q_L^{i \alpha \ T} C Q^{j \beta}_L]
[Q^{k \rho \ T}_L C L^m_{\ell,L}]
[Q^{n \sigma \ T}_L C L^p_{\ell',L}][u_R^{\gamma \ T} C d_R^\delta ] \ 
\in C^{(pp,np,nn)}_{16}
\label{op21_pp}
\eeq
\beq
{\cal O}^{(pp,np,nn)}_{22} = 
(I_{ss})_{ijkm} 
(T_s)_{\alpha\beta\gamma\delta\rho\sigma}
[Q_L^{i \alpha \ T} C L^j_{\ell,L}]
[Q^{k \beta \ T}_L C L^m_{\ell',L}]
[u^{\gamma \ T}_R C u^\delta_R][d_R^{\rho \ T} C d^\sigma_R ] \ \in 
C^{(pp,np,nn)}_{15} 
\label{op22_pp}
\eeq
\beq
{\cal O}^{(pp,np,nn)}_{23} = 
(I_{ss})_{ijkm} 
(T_s)_{\alpha\beta\gamma\delta\rho\sigma}
[Q_L^{i \alpha \ T} C L^j_{\ell,L}]
[Q^{k \beta \ T}_L C L^m_{\ell',L}]
[u^{\gamma \ T}_R C d^\delta_R][u_R^{\rho \ T} C d^\sigma_R ]  \ \in 
C^{(pp,np,nn)}_{15} 
\label{op23_pp}
\eeq
\beq
{\cal O}^{(pp,np)}_{24} = 
(I_{ssa})_{ijkmnp} 
(T_s)_{\alpha\beta\gamma\delta\rho\sigma}
[Q_L^{i \alpha \ T} C Q^{j \beta}_L]
[Q^{k \gamma \ T}_L C Q^{m \delta}_L]
[Q^{n \sigma \ T}_L C L^p_{\ell,L}][u_R^{\sigma \ T} C \ell'_R]  \ \in 
C^{(pp,np,nn)}_{17} 
\label{op24_pp}
\eeq
\beq
{\cal O}^{(pp,np,nn)}_{25} = 
(I_{ssa})_{ijkmnp} (T_s)_{\alpha\beta\gamma\delta\rho\sigma}
[Q_L^{i \alpha \ T} C Q^{j \beta}_L]
[Q^{k \rho \ T}_L C L^m_{\ell,L}]
[Q^{n \sigma \ T}_L C L^p_{\ell',L}][u_R^{\gamma \ T} C d_R^\delta ] 
\ \in C^{(pp,np,nn)}_{16} 
\label{op25_pp}
\eeq
\beq
{\cal O}^{(pp,np,nn)}_{26} = 
(I_{ssa})_{kmnpij} (T_a)_{\alpha\beta\gamma\delta\rho\sigma}
[Q_L^{i \alpha \ T} C Q^{j \beta}_L]
[Q^{k \rho \ T}_L C L^m_{\ell,L}]
[Q^{n \sigma \ T}_L C L^p_{\ell',L}][u_R^{\gamma \ T} C d_R^\delta ]  
\ \in C^{(pp,np,nn)}_{16}
\label{op26_pp}
\eeq
\beq
{\cal O}^{(pp,np,nn)}_{27} = 
(I_{sss})_{ijkmnp} (T_s)_{\alpha\beta\gamma\delta\rho\sigma}
[Q_L^{i \alpha \ T} C Q^{j \beta}_L]
[Q^{k \rho \ T}_L C L^m_{\ell,L}]
[Q^{n \sigma \ T}_L C L^p_{\ell',L}][u_R^{\gamma \ T} C d_R^\delta ] 
\ \in C^{(pp,np,nn)}_{16} 
\label{op27_pp}
\eeq
and
\beq
{\cal O}^{(pp,np,nn)}_{28} =
\epsilon_{ij}\epsilon_{km}\epsilon_{np}\epsilon_{st}
(T_a)_{\alpha\beta\gamma\delta\rho\sigma}
[Q_L^{i \alpha \ T} C Q_L^{j \beta}][Q_L^{k \gamma \ T} C Q_L^{m \delta}] \ .
[Q_L^{n \rho \ T} C L_{\tau,L}^p][Q_L^{s \sigma \ T} C L^t_{\ell,L}] 
\ \in C^{(pp,np,nn)}_{19} \ . 
\label{o28_pp}
\eeq
\end{widetext}
%

% =====================================================================

\end{appendix}

% ======================================================================

\begin{table}
  \caption{\footnotesize{Structures of classes $C^{(Nm3)}_k$ 
      of operators contributing to $\Delta L=-3$ 
      nucleon decays to trileptons. The first column lists the class number; 
      the second column lists the number $N_d$ of SU(2)$_L$ doublets in the 
      operators in this class; and the third column lists the structure of 
      operators in the class. As in the text, we use the 
      abbreviations $pm3$ for $p \to \ell^+ \bar\nu \bar\nu'$ and 
      $nm3$ for $n \to \bar\nu \bar\nu' \bar\nu''$. 
      The abbreviations used for the fermion fields are 
      $Q = Q_L$, $L=L_L$, $u=u_R$, $d=d_R$, $\ell=\ell_R$, and
      $\nu = \nu_{s,R}$. The primes distinguishing different $\nu$ fields are
      suppressed in the notation.}}
\begin{center}
\begin{tabular}{|c|c|c|} \hline\hline
class $C^{(Nm3)}_k$ & $N_d$ & structure \\
\hline
$C^{(pm3)}_1$    & 0 & $u^2 d \ell \nu^2$    \\
$C^{(nm3)}_2$    & 0 & $u d^2 \nu^3$         \\
\hline
$C^{(pm3)}_3$    & 2 & $Q^2 u \ell \nu^2$    \\
$C^{(nm3)}_4$    & 2 & $Q^2 d \nu^3$          \\
$C^{(pm3,nm3)}_5$ & 2 & $Q L u d \nu^2$       \\
$C^{(pm3,nm3)}_6$ & 2 & $Q L u^2 \ell \nu$    \\
\hline
$C^{(pm3,nm3)}_7$ & 4 & $Q^3 L \nu^2$         \\
$C^{(pm3,nm3)}_8$ & 4 & $Q^2 L^2 u \nu$       \\
\hline\hline
\end{tabular}
\end{center}
\label{nuc3_class_table}
\end{table}
%

% =======================================================================

\begin{table}
  \caption{\footnotesize{Structures of classes $C^{(N1)}_k$ 
      of operators contributing to $\Delta L=1$ 
      nucleon decays to trileptons. The first column lists the class number; 
      the second column lists the number $N_d$ of SU(2)$_L$ doublets in the 
      operators in this class; and the third column lists the structure of 
      operators in the class. As in the text, we use the 
      abbreviations $p1$ for $p \to \ell^+ \nu \nu'$ and 
      $n1$ for $n \to \bar\nu \nu' \nu''$. The abbreviations for fermion 
      fields are the same as in Table \ref{nuc3_class_table}. 
      The primes distinguishing different $\nu$ fields are
      suppressed in the notation.}}
\begin{center}
\begin{tabular}{|c|c|c|} \hline\hline
class $C^{(N1)}_k$ & $N_d$ & structure \\
\hline
$C^{(p1)}_1$    & 0 & $u^2 d \ell \bar\nu^2$  \\
$C^{(n1)}_2$    & 0 & $u d^2 \nu \bar\nu^2$   \\
\hline
$C^{(p1)}_3$    & 2 & $Q^2 u \ell \bar\nu^2$  \\
$C^{(n1)}_4$    & 2 & $Q^2 d \nu \bar\nu^2$   \\
$C^{(p1,n1)}_5$ & 2 & $Q L u d \bar\nu^2$     \\
\hline\hline
\end{tabular}
\end{center}
\label{nuc3b_class_table}
\end{table}
%

% ========================================================================

\begin{table}
  \caption{\footnotesize{Structures of classes $C^{(NN')}_k$ 
      of operators contributing to 
      dinucleon-to-dilepton decays with $\Delta L=-2$. 
      The first column lists the class number; the
      second column lists the number of SU(2)$_L$ doublets in the operators in
      this class; and the third 
      column lists the structure of operators in the class. The abbreviations
      in the superscripts on the classes are 
      $pp$ for $pp \to \ell^+ \ell'^+$, $np$ for $np \to \ell^+
      \bar\nu$, and $nn$ for $nn \to \bar\nu \bar\nu'$. 
      The abbreviations for fermion fields are the same as in 
      Table \ref{nuc3_class_table}. The primes distinguishing different lepton
      fields are suppressed in the notation.}}
\begin{center}
\begin{tabular}{|c|c|c|} \hline\hline
class $C^{(NN')}_k$ & $N_d$ & structure  \\
\hline
$C^{(pp)}_1$     & 0 & $u^4 d^2 \ell^2$         \\
$C^{(np)}_2$     & 0 & $u^3 d^3 \ell \nu$       \\
$C^{(nn)}_3$     & 0 & $u^2 d^4 \nu^2$          \\
\hline 
$C^{(pp)}_4$     & 2 & $Q^2 u^3 d \ell^2$        \\
$C^{(np)}_5$     & 2 & $Q^2 u^2 d^2 \ell \nu$    \\
$C^{(nn)}_6$     & 2 & $Q^2 u d^3 \nu^2$         \\
$C^{(pp,np)}_7$     & 2 & $Q L u^3 d^2 \ell$     \\
$C^{(np,nn)}_8$     & 2 & $Q L u^2 d^3 \nu$      \\
$C^{(np)}_9$     & 2 & $L^2 u^3 d^3$             \\
\hline 
$C^{(pp)}_{10}$  & 4 & $Q^4 u^2 \ell^2$          \\
$C^{(np)}_{11}$  & 4 & $Q^4 u d \ell \nu$        \\
$C^{(nn)}_{12}$  & 4 & $Q^4 d^2 \nu^2$           \\
$C^{(pp,np)}_{13}$  & 4 & $Q^3 L u^2 d \ell$     \\
$C^{(np,nn)}_{14}$  & 4 & $Q^3 L u d^2 \nu$      \\
$C^{(pp,np,nn)}_{15}$  & 4 & $Q^2 L^2 u^2 d^2$   \\
\hline 
$C^{(pp,np,nn)}_{16}$  & 6 & $Q^4 L^2 u d$       \\
$C^{(pp,np)}_{17}$  & 6 & $Q^5 L u \ell$         \\
$C^{(np,nn)}_{18}$  & 6 & $Q^5L d \nu$           \\
\hline 
$C^{(pp,np,nn)}_{19}$  & 8 & $Q^6 L^2$           \\
\hline\hline
\end{tabular}
\end{center}
\label{operator_class_table}
\end{table}
%

% ========================================================================

% ======================================================================


\begin{thebibliography}{99}
\bibitem{c1}
%
A different class of models features extra dimensions in which SM
fields propagate but without strong localization of fermion wavefunctions 
\cite{acd}. In the model of \cite{acd}, special symmetries
specific to a particular compactification suppress both proton decay and 
$n-\bar n$ oscillations.
\bibitem{compare_ms}
%
These values are quite close to the values 
$\|\eta_{Q_L}-\eta_{u_R}\|=4.87$ and 
$\|\eta_{Q_L}-\eta_{d_R}\|=4.60$ used in \cite{ms}. The resultant ratio 
$\|\eta_{Q_L}-\eta_{u_R}\|/\|\eta_{Q_L}-\eta_{d_R}\|=1.03$ is the same as 
we get.
\end{thebibliography}
\end{document}